\documentclass[epj]{svjour}
\usepackage{graphics}
\usepackage{amsmath,amssymb,epsfig}
\usepackage{wrapfig}

\begin{document}
\title{Tests of Relativistic Gravity in Space}
\subtitle{Recent Progress and Future Directions}
\author{Slava G. Turyshev\inst{1}\inst{,2} 
}                     
%
%
\institute{Jet Propulsion Laboratory, California Institute of Technology,
4800 Oak Grove Drive, Pasadena, CA 91009 USA
\and Sternberg Astronomical Institute, 13 Universitetskij Prospect, 119992 Moscow, Russia}
\date{Received: date / Revised version: date}
%
\abstract{
We review the foundations of Einstein's general theory of relativity, discuss recent progress in the tests of relativistic gravity, and present motivations for new generation of high-accuracy gravitational experiments.  We discuss the advances in our understanding of fundamental physics anticipated in the near future and evaluate discovery potential of the recently proposed gravitational experiments.
} 
\maketitle

\section{\label{sec:intro}Introduction}

November 25, 2015 will mark the Centennial of general theory of relativity developed by Albert Einstein during the years from 1905 to 1915 \cite{Einstein-1915,Einstein-1916}.  Ever since its original publication, the theory continues to be an active area of research, both theoretical and experimental \cite{Turyshev-etal-2007}.

The theory began with its empirical success in 1915 by explaining the anomalous perihelion precession of Mercury's orbit.  The anomaly was known long before Einstein, it amounts to 43 arcseconds per century (''/cy) and cannot be explained within Newton's gravity, thereby presenting a challenge for physicists and astronomers.  In 1855, Urbain LeVerrier, who predicted the existence of Neptune in 1846, a planet on an extreme orbit, thought that the anomalous residue of the Mercurial precession would be accounted for if yet another planet, still undiscovered planet Vulcan, revolves inside the Mercurial orbit; because of the proximity of the sun it would not be easily observed, but LeVerrier thought he had detected it.  However, no confirmation came in the decades that followed. It took another 60 years to solve this puzzle. In 1915, before publishing the historical paper with the field equations of general relativity (e.g. \cite{Einstein-1915}), Einstein computed the expected perihelion precession of Mercury's orbit; when he got the famous 43 ''/cy needed to account for the anomaly, he realized that a new era in gravitational physics just began!  

Shortly thereafter, Sir Arthur Eddington's 1919 observations of star lines-of-sight during a solar eclipse \cite{Dyson-Eddington-Davidson-1920} confirmed the doubling of the deflection angles predicted by general relativity as compared to Newtonian and EP arguments. Observations were made simultaneously in the city of Sobral, Cear\'a, Brazil and on the island of Principe off the west coast of Africa and focused on determining the change in position of stars as they passed near the Sun on the celestial sphere.  The results were presented November 6, 1919 at a special join meeting of the Royal Astronomical Society and the Royal Society of London \cite{Coles-2001}.  The data from Sorbral, with measurements of seven stars in good visibility, gave the deflections of $1.98\pm0.16$~arcsec.  Principe was less convincing.  Only five stars were included, and the conditions there led to a much larger error.  Nevertheless, the obtained value was $1.61\pm0.4$.  Both were within $2\sigma$ of the Einstein's value of 1.74 and more than two standard deviations away from either zero or the Newtonian value of 0.87. Eddington's 1919 observations became the first dedicated experiment to test general theory of relativity.\footnote{The early accuracy, however, was poor.  Dyson et al. \cite{Dyson-Eddington-Davidson-1920} quoted an optimistically low uncertainty in their measurement, which is argued by some to have been plagued by systematic error and possibly confirmation bias, although modern re-analysis suggests that Eddington's analysis was accurate \cite{Kennefick-2007}.} 
In Europe, still recovering from the World War I, this result was considered spectacular news and occupied the front pages of most major newspapers making general relativity an instant success. 

From these beginnings, general theory of relativity has been verified at ever higher accuracy and presently accounts for all data gathered to date. The true renaissance in the tests of general relativity in the solar system began in 1970s with major advances in microwave spacecraft tracking, high precision astrometric observations, and  lunar laser ranging (LLR). 
Thus, microwave ranging to the Viking Lander on Mars in 1976 yielded a $\sim$0.2\% accuracy in the tests of general relativity \cite{viking_shapiro1,viking_shapiro2}.  Spacecraft and planetary radar observations reached an accuracy of $\sim$0.15\% \cite{Anderson_etal_02,Pitjeva-2005}.  The astrometric observations of quasars on the solar background performed with Very-Long Baseline Interferometry (VLBI) improved the accuracy of the tests of gravity to $\sim$0.045\% \cite{Lebach95,Shapiro_SS_etal_2004}. LLR,  a continuing legacy of the Apollo program, provided $\sim$0.011\% verification of general relativity via precision measurements of the lunar orbit \cite{Williams-etal-2004,Williams-etal-2005}.  Finally, the 2003 microwave tracking of the Cassini spacecraft on its approach to Saturn improved the accuracy of the tests to $\sim$0.0023\% \cite{Iess_etal_1999,Bertotti-Iess-Tortora-2003}. 

To date, general relativity is also in agreement with the data from the binary pulsars.  In fact, recently a considerable interest has been shown in the physical processes occurring in the strong gravitational field regime with relativistic pulsars providing a promising possibility to test gravity in this qualitatively different dynamical environment. The general theoretical framework for pulsar tests of strong-field gravity was introduced in \cite{DamourTaylor92}; the observational data for the initial tests were obtained with PSR1534 \cite{Taylor_etal92}. An analysis of strong-field gravitational tests and their theoretical justification was presented in \cite{Damour-EspFarese-1996-2}.  The recent analysis of the pulsar data tested general relativity to $\sim 0.04\%$ at a $3 \sigma$ confidence level \cite{Lange_etal_2001}. 

It is remarkable that even after over ninety years since general relativity was born, the Einstein's theory has survived every test \cite{Will-lrr-2006-3}.  Such a longevity and success made general relativity the de-facto ``standard'' theory of gravitation for all practical purposes involving spacecraft navigation and astrometry, and also for astronomy, astrophysics, cosmology and fundamental physics \cite{Turyshev-etal-2007}. 
However, despite a remarkable success, there are many important reasons to question the validity of general relativity and to determine the level of accuracy at which it is violated.  

On the theoretical front, the problems arise from several directions, most dealing with the strong gravitational field regime; this includes the appearance of spacetime singularities and the inability of classical description to describe the physics of very strong gravitational fields. A way out of this difficulty would be attained through gravity quantization. However, despite the success of modern gauge field theories in describing the electromagnetic, weak, and strong interactions, it is still not understood how gravity should be described at the quantum level. 

The continued inability to merge gravity with quantum mechanics, and recent cosmological observations indicate that the pure tensor gravity of general relativity needs modification.  In theories that attempt to include gravity, new long-range forces can arise as an addition to the Newtonian inverse-square law. 
Regardless of whether the cosmological constant should be included, there are also important reasons to consider additional fields, especially scalar fields.
Although the latter naturally appear in these modern theories, their inclusion predicts a non-Einsteinian behavior of gravitating systems. These deviations from general relativity lead to a violation of the Equivalence Principle, 
modification of large-scale gravitational phenomena, and cast doubt upon the constancy of the fundamental ``constants.'' These predictions motivate new searches for very small deviations of relativistic gravity from the behavior prescribed by  general relativity; they also provide a new theoretical paradigm and guidance for further gravity experiments \cite{Turyshev-etal-2007}.

In this paper we discuss recent gravitational experiments that contributed to the progress in relativistic gravity research by providing important guidance in the search for the next theory of gravity.  We also present theoretical motivation and proposals for new generation of gravitational experiments.

The paper is organized as follows: Section~\ref{sec:gr-survey} discusses the foundations of general theory of relativity and presents the results of the recent experiments designed to test the  foundations of this theory. Section~\ref{sec:beyond} presents  motivations for extending the theoretical model of gravity provided  by general relativity; it presents models arising from string theory,  discusses the scalar-tensor theories of gravity, and highlights phenomenological implications of these proposals. We also review recent proposals to modify gravity on large scales and addresses the motivations and experimental search for new interactions of nature. Section \ref{sec:future-tests} discusses future space-based experiments aiming to expand our knowledge of gravity. We conclude in Section \ref{sec:conclusions}.

\section{Testing Foundations of General Relativity}
\label{sec:gr-survey}

General relativity is a tensor field theory of gravity with universal coupling to the particles and fields of the Standard Model.  It describes gravity as a universal deformation of the Minkowski metric tenor, $\gamma_{mn}$:
\begin{equation}
g_{mn}(x^k)=\gamma_{mn}+h_{mn}(x^k).
\end{equation}
Alternatively, it can also be defined as the unique, consistent, local theory of a massless spin-2 field $h_{mn}$, whose source is the total, conserved energy-momentum tensor (see \cite{Damour-2006-pdg} and references therein).  

Classically (see \cite{Einstein-1915,Einstein-1916}), general theory of relativity is defined by two postulates.  One of the postulates states that the Lagrangian density describing the propagation and self-inter\-action of the gravitational field is given by
\begin{eqnarray}
\label{eq:hilb-ens-action}
{\cal L}_{\rm G}[g_{mn}]&=&
\frac{c^4}{16\pi G}\sqrt{-g} R,
\end{eqnarray}
where $G$ is the Newton's universal gravitational constant, $g=\det g_{mn}$, and $R$ is the Ricci scalar.  

The second postulate states that $g_{mn}$ couples universally, and minimally, to all the fields of the Standard Model by replacing everywhere the Minkowski metric. Applying the variational principle to the total action 
\begin{equation}
S_{\rm tot}[g_{mn},\psi,A_m,H]=\frac{1}{c}\int d^4x({\cal L}_{\rm G}+{\cal L}_{\rm SM}), 
\end{equation}
one obtains the well-known Einstein's field equations of general theory of relativity,
\begin{equation}
R_{mn}- \frac{1}{2} g_{mn} R= \frac{8\pi G_N}{c^4}T_{mn},
\label{eq:Einstein-eq}
\end{equation}
where $T_{mn}=g_{mk}g_{nl}T^{kl}$ with $T^{mn}=2/\sqrt{-g}~\delta {\cal L}_{\rm SM}/\delta g_{mn}$ being the (symmetric) energy-momentum tensor of the matter as described by the Standard Model. 

In the weak-field and slow motion approximation \cite{Turyshev:1996}, which is sufficient for most of the gravitational experiments in the solar system, Eqs.~(\ref{eq:Einstein-eq}) yield the metric tensor for a system of $N$ point-like gravitational sources:
{}
\begin{eqnarray}
\label{eq:metric}
g_{00}^{\rm GR}&=&1-\frac{2}{c^2}\sum_{j\not=i}\frac{\mu_j}{r_{ij}}+
\frac{2}{c^4}\Big[\sum_{j\not=i}\frac{\mu_j}{r_{ij}}\Big]^2-
\frac{3}{c^4}\sum_{j\not=i}\frac{\mu_j{\dot r}^2_j}{r_{ij}}+\nonumber\\
&+&\frac{2}{c^4}\sum_{j\not=i}\frac{\mu_j}{r_{ij}}
\sum_{k\not=j}\frac{\mu_k}{r_{jk}}-
\frac{1}{c^4}\sum_{j\not=i}\mu_j\frac{\partial^2 r_{ij}}{\partial t^2}+
{\cal O}(c^{-5}),\nonumber\\
g_{0\alpha}^{\rm GR}&=& \frac{4}{c^3}\sum_{j\not=i}\frac{\mu_j{\dot {\bf r}}^\alpha_j}{r_{ij}}+
{\cal O}(c^{-5}),\\\nonumber
g_{\alpha\beta}^{\rm GR}&=&-\delta_{\alpha\beta}\Big(1+\frac{2}{c^2}\sum_{j\not=i}\frac{\mu_j}{r_{ij}}
\Big)+{\cal O}(c^{-5}),   
\end{eqnarray}
\noindent where the indices $j$ and $k$ refer to the $N$ bodies and $k$ includes body $i$, whose motion is being investigated. Also, $\mu_j$ is the gravitational constant for body $j$ given as $\mu_j=Gm_j$, where $G$ is the universal Newtonian gravitational constant and $m_j$ is the isolated rest mass of a body $j$.
In addition, the vector ${\bf r}_i$ is the barycentric radius-vector of this body, the vector ${\bf r}_{ij}={\bf r}_j-{\bf r}_i$ is the
vector directed from body $i$ to body  $j$, $r_{ij}=|{\bf r}_j-{\bf r}_i|$, and the vector ${\bf n}_{ij}={\bf r}_{ij}/r_{ij}$ is the unit vector along this direction. The $1/c^2$ term in $g_{00}$ is the Newtonian limit; the $1/c^4$ terms are post-Newtonian terms. 

The main properties of the metric tensor Eqs.~(\ref{eq:metric}) are well established and widely in use in modern astronomical practice \cite{Turyshev:1996,Moyer-1981-2,Standish_etal_92,Will_book93}
One can derive the Lagrangian function of N-body gravitating system \cite{Turyshev:1996,Will_book93} which is then used to derives the equations of motion for gravitating bodies and light. 
The point-mass Newtonian and relativistic perturbative accelerations in the solar system's barycentric frame can be written in the following form (e.g., the Einstein-Infeld-Hoffmann (EIH) equations):
{}
\begin{eqnarray}
\ddot{\bf r}_i^{\rm GR}&=&\sum_{j\not=i}\frac{\mu_j({\bf r}_j-{\bf r}_i)}{r_{ij}^3}\bigg\{
1-\frac{4}{c^2}\sum_{l\not=i}\frac{\mu_l}{r_{il}}-
\frac{1}{c^2}\sum_{k\not=j}\frac{\mu_k}{r_{jk}}+\nonumber\\
&&+~\big(\frac{{\dot r}_i}{c}\big)^2+2\big(\frac{{\dot r}_j}{c}\big)^2-\frac{4}{c^2} \dot{\bf r}_i \dot{\bf r}_j-\nonumber\\
&&-~\frac{3}{2c^2}\bigg[\frac{({\bf r}_i-{\bf r}_j){\dot{\bf r}}_j}{r_{ij}}\bigg]^2+\frac{1}{2c^2}({\bf r}_j-{\bf r}_i){\ddot{\bf r}}_j\bigg\}\nonumber\\
&&+~\frac{1}{c^2}\sum_{j\not=i}\frac{\mu_j}{r_{ij}^3}
\Big\{\Big[{\bf r}_i-{\bf r}_j\Big]\cdot\Big[ 4{\dot {\bf r}}_i-3{\dot {\bf r}}_j\Big]\Big\}({\dot{\bf r}}_i-{\dot{\bf r}}_j)
\nonumber\\
&&+~\frac{7}{2c^2}\sum_{j\not=i}\frac{\mu_j{\ddot {\bf r}}_j}{r_{ij}}+{\cal O}(c^{-4}).
\label{eq:4-26-mod}
\end{eqnarray}
\noindent 
The general relativistic equations of motion Eq.~(\ref{eq:4-26-mod}) are then used to produce numerical codes for the purposes of construction solar system ephemerides, spacecraft orbit determination \cite{Moyer-2003,Standish_etal_92,Turyshev:1996}, and analysis of the gravitational experiments in the solar system \cite{Turyshev-etal-2007,Will_book93,Turyshev_etal_acfc_2003}.

\subsection{Scalar-Tensor Extensions to General Relativity}
\label{sec:scalar-tensor}

Among the theories of gravity alternative to general relativity, metric theories have a special place among other possible theoretical models. The reason is that, independently of the different principles at their foundations, the gravitational field in these theories affects the matter directly through the metric tensor $g_{mn}$, which is determined from the field equations of a particular theory. As a result, in contrast to Newtonian gravity, this tensor expresses the properties of a particular gravitational theory and carries information about the gravitational field of the bodies.

In many alternative theories of gravity, the gravitational coupling strength exhibits a dependence on a field of some sort; in scalar-tensor theories, this is a scalar field $\varphi$. A general action for these theories can be written as
{}
\begin{eqnarray}  
S&=&{c^3\over 4\pi G}\int d^4x \sqrt{-g}
\left[\frac{1}{4}f(\varphi) R - \frac{1}{2}g(\varphi) \partial_{\mu} \varphi
\partial^{\mu} \varphi + V(\varphi) \right] \nonumber\\
&+& \sum_{i}
q_{i}(\varphi)\mathcal{L}_{i}, \label{eq:sc-tensor} 
\end{eqnarray}
\noindent where $f(\varphi)$, $g(\varphi)$, $V(\varphi)$ are generic functions, $q_i(\varphi)$ are coupling functions and $\mathcal{L}_{i}$ is the Lagrangian density of the matter fields. The well-known Brans-Dicke theory \cite{Brans-Dicke-1961} corresponds to the specific choice
{}
\begin{equation} 
\label{eq6:2} 
f(\varphi) = \varphi, \qquad g(\varphi) =
{\omega \over \varphi}, \qquad V(\varphi)=0.
\end{equation}

Notice that in the Brans-Dicke theory the kinetic energy term of the field $\varphi$ is non-canonical, and the latter has a dimension of energy squared. In this theory, the constant $\omega$ marks observational deviations from general relativity, which is recovered in the limit $\omega \to \infty$. We point out that, in the context of the Brans-Dicke theory, one can operationally introduce the Mach's
Principle which, we recall, states that the inertia of bodies is due to their interaction with the matter distribution in the Universe. Indeed, in this theory the gravitational coupling is proportional to $\varphi^{-1}$, which depends on the energy-momentum tensor of matter through the field equations. 
Observational bounds require that $|\omega| \gtrsim 40000$ \cite{Bertotti-Iess-Tortora-2003,Will-lrr-2006-3}. Other alternative theories exist and are used to provide guidance for gravitational experiments (see details in \cite{Will-lrr-2006-3}).

\subsection{PPN Extension of General Relativity}
\label{sec:ppn}

Generalizing on a phenomenological parameterization of the gravitational metric tensor field, which Eddington originally developed for a special case, a method called the parameterized post-Newtonian (PPN) formalism has been developed \cite{Will_book93}. This method represents the gravity tensor's potentials for slowly moving bodies and weak inter-body gravity, and is valid for a broad class of metric theories, including general relativity as a unique case. The several parameters in the PPN metric expansion vary from theory to theory, and they are individually associated with various symmetries and invariance properties of the underlying theory (for more details, consult \cite{Will_book93}). 

In a special case (e.g., if, one assumes that Lorentz invariance, local position invariance and total momentum conservation hold), when only two PPN parameters ($\gamma$, $\beta$) are considered, these parameters have clear physical meaning. The parameter $\gamma$  represents the measure of the curvature of the space-time created by a unit rest mass; parameter  $\beta$ is a measure of the non-linearity of the law of superposition of the gravitational fields in the theory of gravity. General relativity, when analyzed in standard PPN gauge, gives: $\gamma=\beta=1$ and all the other eight parameters vanish; the theory is thus embedded in a two-dimensional space of theories. 
 
The Brans-Dicke theory \cite{Brans-Dicke-1961} is the best known of the alternative theories of gravity. It contains, besides the metric tensor, a scalar field and an arbitrary coupling constant $\omega$, which yields the two PPN parameter values, $\beta=1$, $\gamma= ( 1 + \omega ) / ( 2 + \omega )$, where $\omega$ is an unknown
dimensionless parameter of this theory. More general scalar tensor theories yield values of $\beta$ different from one \cite{Turyshev:1996,Damour_Nordtvedt_93b}.

In the complete PPN framework, a particular metric theory of gravity in the PPN formalism with a specific coordinate gauge is fully characterized by means of ten PPN parameters \cite{Turyshev:1996,Will_book93}. Thus, besides the parameters $\gamma, \beta$, there other eight parameters $\alpha_1, \alpha_2, \alpha_3, \zeta, \zeta_1,\zeta_2,\zeta_3,\zeta_4$. The formalism uniquely prescribes the values of these parameters for each particular theory under study. Gravity experiments can be analyzed in terms of the PPN metric, and an ensemble of experiments will determine the unique value for these parameters, and hence the metric field itself. 

Given the phenomenological success of  general relativity, it is convenient to use this theory to describe the experiments. In this sense, any possible deviation from general relativity would represent itself as a small perturbation to this general relativistic background. These perturbations are proportional to re-normalized PPN parameters (i.e., $\bar{\gamma}\equiv\gamma-1,\bar{\beta}\equiv\beta-1$, etc.) that are zero in general relativity, but may have non-zero values for some gravitational theories. In terms of the metric tensor, this PPN-perturbative procedure may be given as 
{}
\begin{equation}
\label{eq:metric2}
g_{mn}=g_{mn}^{\rm GR}+\delta g^{\rm PPN}_{mn},  
\end{equation}
where metric $g_{mn}^{\rm GR}$ is derived from Eq.~(\ref{eq:metric}) by taking the general relativistic values of the PPN parameters. 

If, one assumes that Lorentz invariance, local position invariance and total momentum conservation hold,  the PPN-renormalized metric perturbation $\delta g^{\rm PPN}_{mn}$ for a system of $N$ point-like gravitational sources in four dimensions may be given as 
\begin{eqnarray}
\label{eq:metric-PPN}
\delta g^{\rm PPN}_{00}&=&-
\frac{2{\bar \gamma}}{c^4}\sum_{j\not=i}\frac{\mu_j{\dot r}^2_j}{r_{ij}}+\frac{2{\bar \beta}}{c^4}\Big(\Big[\sum_{j\not=i}\frac{\mu_j}{r_{ij}}\Big]^2+\nonumber\\
&+&2\sum_{j\not=i}\frac{\mu_j}{r_{ij}}\sum_{k\not=j}\frac{\mu_k}{r_{jk}}\Big)+
{\cal O}(c^{-5}),\nonumber\\
\delta g^{\rm PPN}_{0\alpha}&=& \frac{2{\bar \gamma}}{c^3}\sum_{j\not=i}\frac{\mu_j{\dot {\bf r}}^\alpha_j}{r_{ij}}+
{\cal O}(c^{-5}),\\\nonumber
\delta g^{\rm PPN}_{\alpha\beta}&=&-\delta_{\alpha\beta}\frac{2{\bar \gamma}}{c^2}\sum_{j\not=i}\frac{\mu_j}{r_{ij}}
+{\cal O}(c^{-5}).  
\end{eqnarray}
\noindent Given the smallness of the current values for the PPN parameters $\bar{\gamma}$ and $\bar{\beta}$, the PPN metric perturbation $\delta g^{\rm PPN}_{mn}$ represents a very small deformation of the general relativistic background $g_{mn}^{\rm GR}$.  Expressions Eqs.~(\ref{eq:metric-PPN}) represent the `spirit' of many tests  when one essentially assumes that general relativity provides correct description of the experimental situation and searches for small deviations. 

\begin{figure}[h!]
  \vspace{-15pt}
  \begin{center}
    \includegraphics[width=0.46\textwidth]{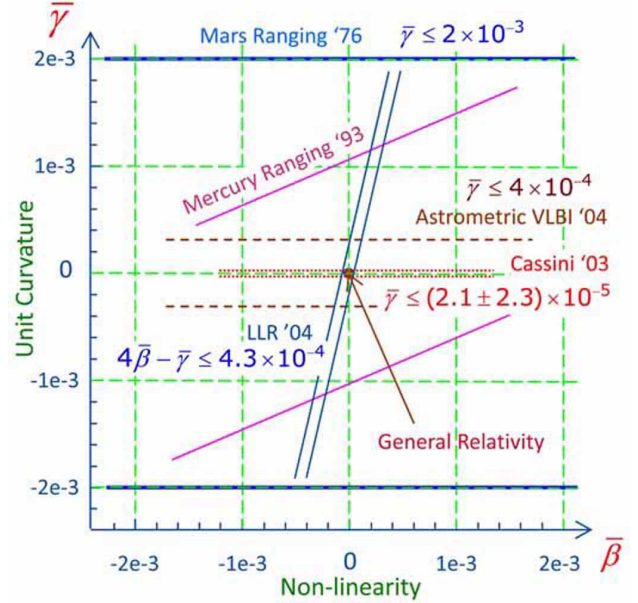}
  \end{center}
  \vspace{-10pt}
  \caption{The progress in improving the knowledge of the PPN parameters $\gamma$ and $\beta$ for the last 39 years (i.e., since 1969 \cite{Williams-etal-2005}). So far, general theory of relativity survived every test \cite{Turyshev-etal-2007}, yielding $\bar\gamma=(2.1\pm2.3)\times10^{-5}$ \cite{Bertotti-Iess-Tortora-2003} and $\bar\beta=1.1\times10^{-4}$ \cite{Williams-etal-2004}.}
\label{fig:gamma-beta}
  \vspace{-10pt}
\end{figure}

Similarly, one derives the PPN-renormalized the equations of motion Eq.~(\ref{eq:4-26-mod}) may also be presented with explicit dependence on the PPN perturbative acceleration terms:
\begin{equation}
\label{eq:eq-m2}
\ddot{\bf r}_i^{\rm PPN}=\ddot{\bf r}_i^{\rm GR}+\delta\ddot{\bf r}_i^{\rm PPN},  
\end{equation}
with  $\ddot{\bf r}_i^{\rm GR}$ being the equations of motion Eq.~(\ref{eq:4-26-mod}), while the PPN perturbative acceleration term $\delta\ddot{\bf r}_i^{\rm PPN}$ is given as
{}
\begin{eqnarray}
\delta\ddot{\bf r}_i^{\rm PPN}&=&\sum_{j\not=i}\frac{\mu_j({\bf r}_j-{\bf r}_i)}{r_{ij}^3}\bigg\{
 \Big(\left[\frac{m_G}{m_I}\right]_i-1\Big)
+\frac{{\dot G}}{G}\cdot (t-t_0)-\nonumber\\
&-&
\frac{2({\bar \beta}+{\bar\gamma})}{c^2}\sum_{l\not=i}\frac{\mu_l}{r_{il}}-
\frac{2{\bar\beta}}{c^2}\sum_{k\not=j}\frac{\mu_k}{r_{jk}}+\frac{{\bar\gamma}}{c^2} (\dot{\bf r}_i-\dot{\bf r}_j)^2\bigg\}\nonumber\\
&+&\frac{2{\bar\gamma}}{c^2}\sum_{j\not=i}\frac{\mu_j}{r_{ij}^3}
\Big\{\big({\bf r}_i-{\bf r}_j\big)\cdot\big( {\dot {\bf r}}_i-{\dot {\bf r}}_j\big)\Big\}({\dot{\bf r}}_i-{\dot{\bf r}}_j)+\nonumber\\
&+&\frac{2{\bar\gamma}}{c^2}\sum_{j\not=i}\frac{\mu_j{\ddot {\bf r}}_j}{r_{ij}}+{\cal O}(c^{-4}).
\label{eq:4-26-mod-PPN}
\end{eqnarray}
\noindent Eq.~(\ref{eq:4-26-mod-PPN}) provides a useful framework for gravitational research. Thus, besides the terms with PPN-renormalized parameters $\bar\gamma$ and $\bar\beta$, it also contains $([{m_G}/{m_I}]_i-1)$, the parameter that signifies possible inequality of the gravitational and inertial masses and is needed to facilitate investigation of a possible violation of the  Equivalence Principle (see Sec.~\ref{sec:sep}); in addition, Eq.~(\ref{eq:4-26-mod-PPN}) also includes parameter ${\dot G}/{G}$, needed to investigate possible temporal variation in the gravitational constant (see Sec.~\ref{sec:variation-G}).

Eqs.~(\ref{eq:eq-m2}) and (\ref{eq:4-26-mod-PPN}) are used to focus the science objectives and to describe gravitational experiments (especially those to be conducted in the solar system) that well be discussed below. So far, general theory of relativity survived every test \cite{Turyshev-etal-2007}, yielding the ever improving values for the PPN parameters $(\gamma,\beta)$, namely $\bar\gamma=(2.1\pm2.3)\times10^{-5}$ using the data from the Cassini spacecraft taken during solar conjunction experiment \cite{Bertotti-Iess-Tortora-2003} and $\bar\beta=1.1\times10^{-4}$ resulted from the analysis the LLR data \cite{Williams-etal-2004} (see Fig.~\ref{fig:gamma-beta}).

\section{Search for New Physics Beyond General Relativity}
\label{sec:beyond}

The fundamental physical laws of Nature, as we know them today, are described by the Standard Model of particles and fields and general theory of relativity. 
The Standard Model specifies the families of fermions (leptons and quarks) and their interactions by vector fields which transmit the strong, electromagnetic, and weak forces. General relativity is a tensor field theory of gravity with universal coupling to the particles and fields of the Standard Model.  

However, despite the beauty and simplicity of general relativity and the success of the Standard Model, our present understanding of the fundamental laws of physics has several shortcomings. Although recent  progress in string theory \cite{Witten:2001,Witten:2003} is very encouraging,  the search for a realistic theory of quantum gravity remains a challenge. This continued inability to merge gravity with quantum mechanics indicates that the pure tensor gravity of general relativity needs modification or augmentation. 
The recent remarkable progress in observational cosmology has subjected general theory of relativity to increased scrutiny by suggesting a non-Einsteinian model of the universe's evolution.  It is now believed that new physics is needed to resolve these issues.

Theoretical models of the kinds of new physics that can solve the problems above typically involve new interactions, some of which could manifest themselves as violations of the equivalence principle, variatixon of fundamental constants, modification of the inverse square law of gravity at short distances, Lorenz symmetry breaking, as well as large-scale gravitational phenomena.   Each of these manifestations offers an opportunity for space-based experimentation and, hopefully, a major discovery.

Below we discuss motivations for the new generation of gravitational experiments that are expected to advance the relativistic gravity research up to five orders of magnitude below the level currently tested by experiment \cite{Turyshev-etal-2007}.

%
\subsection{String/M-Theory and Tensor-Scalar Extensions of General Relativity}
\label{sec:SMT}

An understanding of gravity at a quantum level will allow one to ascertain whether the gravitational ``constant'' is a running coupling constant like those of other fundamental interactions of Nature. String/M-theory \cite{Green-Schwarz-Witten-1987} hints a negative answer to this question, given the non-renormalization theorems of supersymmetry, a symmetry at the core of the underlying principle of string/M-theory and brane models, \cite{Polchinski-95,Horava:1995qa,Lukas-1999}. 1-loop higher--derivative quantum gravity models may permit a running gravitational coupling, as these models are asymptotically free, a striking property \cite{Fradkin-Tseytlin-1987,Avramidi-Barvinsky-1982}. In the absence of a screening mechanism for gravity, asymptotic freedom may imply that quantum gravitational corrections take effect on
macroscopic and even cosmological scales, which of course has some bearing on the dark matter problem \cite{Goldman-etal-1992} and, in particular, on the subject of the large scale structure of the Universe. Either way, it seems plausible to assume that quantum gravity effects manifest themselves only on cosmological scales.

Description of quantum matter in a classical gravitational background poses interesting challenges, notably the possibility that the zeropoint fluctuations of the matter fields generate a non-vanishing vacuum energy density $\rho_{\rm vac}$, corresponding to a term $-\sqrt{-g}\rho_{\rm vac}$, in ${\cal L}_{\rm SM}$ \cite{Weinberg-1989:RevModPhys}. This is equivalent to adding a ``cosmological constant'' term $+\Lambda g_{mn}$ on the left-hand side of Einstein's equations Eq.~(\ref{eq:Einstein-eq}), with $\Lambda = 8\pi G_M \rho_{\rm vac}/c^4$. Recent cosmological observations suggest a positive value of $\Lambda$ corresponding to $\rho_{\rm vac} \approx (2.3 \times 10^{-3} eV)^4$.  Such a small value has a negligible effect on the solar system dynamics and relevant gravitational tests. Quantizing the gravitational field itself poses a very difficult challenge because of the perturbative non-renormalizability of Einstein's Lagrangian. Superstring theory offers a promising avenue toward solving this challenge.

String theory is viewed as the most promising scheme to make general relativity compatible with quantum mechanics \cite{Green-Schwarz-Witten-1987}. The closed string theory has a spectrum that contains as zero mass eigenstates the graviton, $g_{MN}$, the dilaton, $\Phi$, and an antisymmetric second-order tensor, $B_{MN}$. There are various ways in which to extract the physics of our four-dimensional world, and a major difficulty lies in finding a natural mechanism that fixes the value of the dilaton field, since it does not acquire a potential at any order in string perturbation theory.

Damour and Polyakov \cite{Damour_Polyakov_94b} have studied a possible a mechanism to circumvent the above difficulty by suggesting string loop-contributions, which are counted by dilaton interactions, instead of a potential. After dropping the antisymmetric second-order tensor and introducing fermions, $\hat \psi$, Yang-Mills fields, $\hat A^{\mu}$, with field strength $\hat F_{\mu \nu}$, in a spacetime described by the metric $\hat g_{\mu \nu}$, the relevant effective low-energy four-dimensional
action is

\begin{eqnarray} 
S &=& \int_M d^4 x \sqrt{-\hat g} B(\Phi)
\Big[{1 \over \alpha^{\prime}} [\hat R + 4 \hat \nabla_{\mu} \hat \nabla^{\mu}\Phi - 4 (\hat \nabla \Phi)^2] - \nonumber\\
&&\hskip 60pt-
{k \over 4} \hat F_{\mu \nu} \hat F^{\mu \nu} - \overline{\hat \psi} \gamma^{\mu} \hat D_{\mu} \hat
\psi + ... \Big], \label{eq:3.1} 
\end{eqnarray}
\noindent where
$B(\Phi) = e^{-2 \Phi} + c_0 + c_1 e^{2 \Phi} + c_2 e^{4 \Phi}
+ ..., $
also, $\alpha^{\prime}$ is the inverse of the string tension and $k$ is a gauge group constant; the constants $c_0$, $c_1$, ..., can, in principle, be determined via computation.

To recover Einsteinian gravity, one performs a conformal transformation with $ g_{\mu \nu} = B(\Phi) \hat g_{\mu \nu}$, that leads to an action where the coupling constants and masses are functions of the rescaled dilaton, $\phi$:
{}
\begin{eqnarray} 
S &=& \int_M d^4 x \sqrt{-g} \Big[{1 \over 4q} R - {1 \over
2q} (\nabla \phi)^2 - \nonumber\\
&&\hskip 30pt-
 {k \over 4} B(\phi) F_{\mu \nu} F^{\mu \nu}
- \overline{\psi} \gamma^{\mu} D_{\mu} \psi + ... \Big],
\label{eq:3.4} 
\end{eqnarray}

\noindent from which follows that $4q = 16 \pi G = {1 \over 4} \alpha^{\prime}$ and the coupling constants and masses are now dilaton-dependent, through $ g^{-2} = k B(\phi)$ and $ m_A = m_A(B(\phi))$. 

Damour and Polyakov proposed the minimal coupling
principle (MCP), stating that the dilaton is dynamically driven towards a local minimum of all masses (corresponding to a local maximum of {$B(\phi)$). Due to the MCP, the dependence of the masses on the dilaton implies that particles fall differently in a
gravitational field, and hence are in violation of the WEP. Although, in the solar system conditions, the effect is rather small  being of the order of $\Delta a / a \simeq 10^{-18}$, application of already available technology can potentially test
prediction which represent a distinct experimental signature of string/M-theory. 

\begin{figure}[h!]
  \vspace{-00pt}
  \begin{center}
    \includegraphics[width=0.4\textwidth]{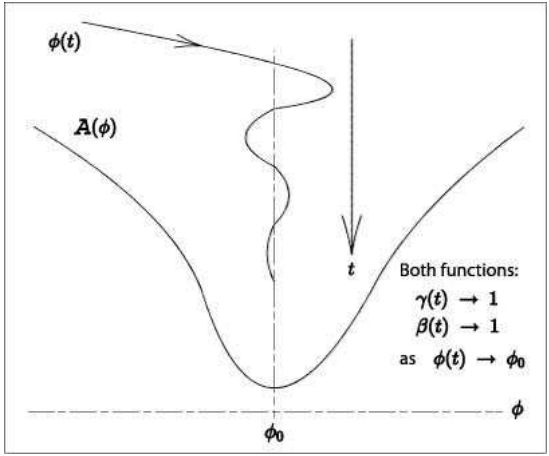}
  \end{center}
  \vspace{0pt}
\caption{\label{fig:attract}
Typical cosmological dynamics of a background scalar field is shown in the case when that field's coupling function to matter, $V(\phi)$, has an attracting point $\phi_0$. The strength of the scalar interaction's coupling to matter is proportional to the derivative (slope) of the coupling function, so it weakens as the attracting point is approached, and both the Eddington parameters $\gamma$ and $\beta$ (and all higher structure parameters as well)  approach their pure tensor gravity values in this limit \cite{Damour-EspFarese-1996-2,Damour_Nordtvedt_93b,Damour_Piazza_Veneziano_02b}.  But a small residual scalar gravity should remain today because this dynamical process is not complete \cite{Turyshev-etal-2006}.
}
  \vspace{-10pt}
\end{figure}

These recent theoretical findings suggest that the present agreement between general relativity and experiment might be naturally compatible with the existence of a scalar contribution to gravity. In particular, Damour and Nordtvedt \cite{Damour_Nordtvedt_93b} (see also \cite{Damour_Polyakov_94b} for non-metric versions of this mechanism together with \cite{Damour_Piazza_Veneziano_02b} for the recent summary of a dilaton-runaway scenario) have found that a scalar-tensor theory of gravity may contain a ``built-in'' cosmological attractor mechanism toward general relativity.  These scenarios assume that the scalar coupling parameter $\frac{1}{2}\bar\gamma$ was of order one in the early universe (say, before inflation), and show that it then evolves to be close to, but not exactly equal to, zero at the present time (Fig.~\ref{fig:attract} illustrates this mechanism in more details). 

The Eddington parameter $\gamma$, whose value in general relativity is unity, is perhaps the most fundamental PPN parameter, in that $\frac{1}{2}\bar\gamma$ is a measure, for example, of the fractional strength of the scalar gravity interaction in scalar-tensor theories of gravity \cite{Damour-EspFarese-1996-2}.  Within perturbation theory for such theories, all other PPN parameters to all relativistic orders collapse to their general relativistic values in proportion to $\frac{1}{2}\bar\gamma$. Under some assumptions (see e.g. \cite{Damour_Nordtvedt_93b}) one can even estimate what is the likely order of magnitude of the left-over coupling strength at present time which, depending on the total mass density of the universe, can be given as $\bar\gamma \sim 7.3 \times 10^{-7}(H_0/\Omega_0^3)^{1/2}$, where $\Omega_0$ is the ratio of the current density to the closure density and $H_0$ is the Hubble constant in units of 100 km/sec/Mpc. Compared to the cosmological constant, these scalar field models are consistent with the supernovae observations for a lower matter density, $\Omega_0\sim 0.2$, and a higher age, $(H_0 t_0) \approx 1$. If this is indeed the case, the level $\bar\gamma \sim 10^{-6}-10^{-7}$ would be the lower bound for the present value of PPN parameter $\bar\gamma$ \cite{Damour_Nordtvedt_93b}. 

Recently, \cite{Damour_Piazza_Veneziano_02b} have estimated $\frac{1}{2}\bar\gamma$, within the framework compatible with string theory and modern cosmology, confirming the previous result \cite{Damour_Nordtvedt_93b}. This recent analysis discusses a scenario when a composition-independent coupling of dilaton to hadronic matter produces detectable deviations from general relativity in high-accuracy light deflection experiments in the solar system. This work assumes only some general property of the coupling functions (for large values of the field, i.e. for an ``attractor at infinity'') and then only assume that $\bar\gamma$ is of order of one at the beginning of the controllably classical part of inflation.
It was shown  in \cite{Damour_Piazza_Veneziano_02b} that one can relate the present value of $\frac{1}{2}\bar\gamma$ to the cosmological density fluctuations. For the simplest inflationary potentials, \cite{Damour_Piazza_Veneziano_02b} found that the present value of $\bar\gamma$ could be just below $10^{-7}$. 

It should be noted that for the run-away dilaton scenario presented here, comparison with the minimally coupled scalar field action,
{}
\begin{equation} S_{\phi} = {c^3\over 4\pi G}\int d^{4}x\sqrt{-g}
\left[\frac{1}{4}R +\frac{1}{2}\partial_{\mu} \phi\partial^{\mu}
\phi-V(\phi)\right],\end{equation}
{}
\noindent reveals that the negative scalar kinetic term leads to an action equivalent to a ``ghost'' in quantum field theory, and is referred to as ``phantom energy'' in the cosmological context \cite{Caldwell:1999ew}. Such a scalar field model could in theory generate acceleration by the field evolving {\it up} the potential toward the maximum. Phantom fields are plagued by catastrophic UV instabilities, as particle excitations have a negative mass \cite{Cline:2003gs,Rubakov:2008nh,Sergienko:2008tf}; the fact that their energy is unbounded from below allows vacuum decay through the production of high energy real particles and negative energy ghosts that will be in contradiction with the constraints on ultra-high energy cosmic rays \cite{Sreekumar_etal:1998}.

Such runaway behavior can potentially be avoided by the introduction of higher-order kinetic terms in the action.  One implementation of this idea is ``ghost condensation'' \cite{ArkaniHamed:2003uz}. Here, the scalar field has a negative kinetic energy near $\dot\phi=0$, but the quantum instabilities are stabilized by the addition of higher-order corrections to the scalar field Lagrangian of the form $(\partial_{\mu} \phi\partial^{\mu} \phi)^{2}$. The ``ghost'' energy is then bounded from below, and stable evolution of the dilaton occurs with $w\ge-1$ \cite{Piazza:2004df}.  The gradient $\partial_\mu\phi$ is non-vanishing in the vacuum, violating Lorentz invariance, and may have important consequences in cosmology and in laboratory experiments.

The analyses discussed above predict very small observable post-Newtonian deviations from general relativity in the solar system in the range from $10^{-5}$ to $5\times10^{-8}$ for $\frac{1}{2}\bar\gamma$, thereby motivating new generation of advanced gravity experiments. This is why measurement of the first order light deflection effect at the level of accuracy comparable with the second-order contribution would provide the crucial information separating alternative scalar-tensor theories of gravity from general relativity and also to probe possible ways for gravity quantization and to test modern theories of cosmological evolution \cite{Damour_Nordtvedt_93b,Damour_Polyakov_94b,Damour_Piazza_Veneziano_02b}. In many cases, such tests would require measurement of the effects of the next post-Newtonian order ($\propto G^2$) \cite{Turyshev-etal-2007,Turyshev-LATOR:2003}, which could lead to important outcomes for the 21st century fundamental physics. 

\subsection{Modified Gravity as an Alternative to Dark Energy}
\label{sec:de}

One can account for dark energy by modifying the Einstein-Hilbert action Eq.~(\ref{eq:hilb-ens-action}) by adding terms that are blow up as the scalar curvature goes to zero \cite{Carroll:2004de}.  Recently, models involving inverse powers of the curvature have been proposed as an alternative to dark energy. In these models one generically has more propagating degrees of freedom in the gravitational sector than the two contained in the massless graviton in general relativity. The simplest models of this kind add inverse powers of the scalar curvature to the action ($\Delta {\cal L}\propto 1/R^n$), thereby introducing a new scalar excitation in the spectrum. For the values of the parameters required to explain the acceleration of the Universe this scalar field is almost massless in vacuum, thus, leading to a possible conflict to data from solar system experiments. 

However, it can be shown that models that involve inverse powers of other invariant, in particular those that diverge for $r\rightarrow 0$ in the Schwarzschild solution, generically recover an acceptable weak field limit at short distances from sources by means of a screening or shielding of the extra degrees of freedom at short distances \cite{Navarro:2005da}. Such theories can lead to late-time acceleration, but typically lead to one of two problems. Either they are in conflict with tests of general relativity in the solar system, due to the existence of additional dynamical degrees
of freedom \cite{Chiba:2003ir}, or they contain ghost-like degrees of freedom that seem difficult to reconcile with fundamental theories.   

A more dramatic approach would be to imagine that we live on a brane embedded in a large extra dimension.  Although such theories can lead to perfectly conventional gravity on large scales, it is also possible to choose the dynamics in such a way that new effects show up exclusively in the far infrared. An example of recent theoretical progress is  the Dvali-Gabadadze-Porrati (DGP) brane-world model, which explores a possibility that  we live on a brane embedded in a large extra dimension, and where the strength of gravity in the bulk is substantially less than that on the brane \cite{Dvali-Gabadadze-Porrati-2003}. Although such theories can lead to perfectly conventional gravity on large scales, it is also possible to choose the dynamics in such a way that new effects show up exclusively in the far infrared providing a mechanism to explain the acceleration of the universe \cite{Perlmutter:1998np,Riess_supernovae98}.  It is interesting to note that DGP gravity and other modifications of general relativity hold out the possibility of having interesting and testable predictions that distinguish them from models of dynamical dark energy. One outcome of this work is that the physics of the accelerating universe may be deeply tied to the properties of gravity on relatively short scales, from millimeters to astronomical units.

Although many effects expected by gravity modification models are suppressed within the solar system, there are measurable effects induced by some long-distance modifications of gravity \cite{Dvali-Gabadadze-Porrati-2003}. For instance, in the case of the precession of the planetary perihelion in the solar system, the anomalous perihelion advance, $\Delta \phi$, induced by a small correction, $\delta U_N$, to Newton's potential, $U_N$, is given in radians per revolution \cite{Dvali-Gruzinov-Zaldarriaga-2003} by
$\Delta \phi \simeq \pi r {d \over dr} [r^2 {d \over dr}({\delta U_N \over rU_N})].$ The most reliable data regarding the planetary perihelion advances come from the inner planets of the solar system, where most of the corrections are negligible. However, LLR offers an interesting possibility to test for these new effects \cite{Williams-etal-2004}. Evaluating the expected magnitude of the effect to the Earth-Moon system, one predicts an anomalous shift of $\Delta \phi \sim 10^{-12}$ \cite{Dvali-Gruzinov-Zaldarriaga-2003}, to be compared with the achieved accuracy of $2.4\times 10^{-11}$. Therefore, the theories of gravity modification result in an intriguing possibility of discovering new physics, if one focuses on achieving higher precision in modern astrometrical measurements; this accuracy increase is within the reach and should be attempted in the near future.

\subsection{Scalar Field Models as Candidate for Dark Energy}
\label{sec:sc-models-de}

One of the simplest candidates for dynamical dark energy is a scalar field, $\varphi$, with an extremely low-mass and an effective potential, $V(\varphi)$.  If the field is rolling slowly, its persistent potential energy is responsible for creating the late epoch of inflation we observe today. For the models that include only inverse powers of the curvature, besides the Einstein-Hilbert term, it is however possible that in regions where the curvature is large the scalar has naturally a large mass and this could make the dynamics to be similar to those of general relativity \cite{Cembranos:2005fi}. At the same time, the scalar curvature, while being larger than its mean cosmological value, it is very small in the solar system thereby satisfying constraints set by  gravitational tests performed to date \cite{Erickcek:2006vf,Nojiri:2006gh}.
Nevertheless, it is not clear whether these models may be regarded as a viable alternative to dark energy.

Effective scalar fields are prevalent in supersymmetric field theories and string/M-theory. For example, string theory predicts that the vacuum expectation value of a scalar field, the dilaton, determines the relationship between the gauge and gravitational couplings. A general, low energy effective action for the massless modes of the dilaton can be cast as a scalar-tensor theory as Eq.~(\ref{eq:sc-tensor}) with a vanishing potential, where $f(\varphi)$, $g(\varphi)$ and $q_{i}(\varphi)$ are the dilatonic couplings to gravity, the scalar kinetic term and gauge and matter fields respectively, encoding the effects of loop effects and potentially non-perturbative corrections.

A string-scale cosmological constant or exponential dilaton potential in the string frame translates into an exponential potential in the Einstein frame. Such quintessence potentials \cite{Wetterich:2004ss,Peebles:2002gy,Wetterich:2004pv} can have scaling \cite{Ferreira_Joyce:1997}, and tracking properties that allow the scalar field energy density to evolve alongside the other matter constituents. A problematic feature of scaling potentials \cite{Ferreira_Joyce:1997} is that they do not lead to accelerative expansion, since the energy density simply scales with that of matter. Alternatively, certain potentials can predict a dark energy density which alternately dominates the Universe and decays away; in such models, the acceleration of the Universe is transient \cite{Albrecht_Skordis:2000,Bento_Bertolami_Santos:2002}. Collectively, quintessence potentials predict that the density of the dark energy dynamically evolve in time, in contrast to the cosmological constant. Similar to a cosmological constant, however, the scalar field is expected to have no significant density perturbations within the causal horizon, so that they contribute little to the evolution of the clustering of matter in large-scale structure \cite{Ferreira_Joyce:1998}.

\section{\label{sec:future-tests} Search for a new theory of gravity with space-based experiments}

It is known that the work on general theory of relativity began with Equivalence Principle (EP), in which gravitational acceleration was held a priori indistinguishable from acceleration caused by mechanical forces; as a consequence, gravitational mass was therefore identical to inertial mass. 
Since Newton, the question about the equality of inertial and passive gravitational masses has risen in almost every theory of gravitation. Einstein elevated this identity, which was implicit in Newton's gravity, to a guiding principle in his attempts to explain both electromagnetic and gravitational acceleration according to the same set of physical laws. Thus, almost one hundred years ago Einstein postulated that not only mechanical laws of motion, but also all non-gravitational laws should behave in freely falling frames as if gravity was absent. It is this principle that predicts identical accelerations of compositionally different objects in the same gravitational field, and also allows gravity to be viewed as a geometrical property of spacetime--leading to the general relativistic interpretation of gravitation.  

Remarkably, but EP has been (and still is!) a focus of gravitational research for more than four hundred years \cite{Williams-etal-2005}.  Since the time of Galileo (1564-1642) it has been known that objects of different mass and composition accelerate at identical rates in the same gravitational field.  In 1602-04 through his study of inclined planes and pendulums, Galileo formulated a law of falling bodies that led to an early empirical version of the EP.  However, these famous results would not be published for another 35 years.  It took an additional fifty years before a theory of gravity that described these and other early gravitational experiments was published by Newton (1642-1727) in his Principia in 1687.  Newton concluded on the basis of his second law that the gravitational force was proportional to the mass of the body on which it acted, and by the third law, that the gravitational force is proportional to the mass of its source. 

Newton was aware that the {\it inertial mass} $m_I$ appearing in the second law ${\bf F} = m_I {\bf a}$, might not be the same as the {\it gravitational mass} $m_G$ relating force to gravitational field  ${\bf F} = m_G {\bf g}$. Indeed, after rearranging the two equations above we find ${\bf a} = (m_G/m_I){\bf g}$ and thus, in principle, materials with different values of the ratio $(m_G/m_I)$ could accelerate at different rates in the same gravitational field.  He went on testing this possibility with simple pendulums of the same length but different masses and compositions, but found no difference in their periods.  On this basis Newton concluded that $(m_G/m_I)$ was constant for all matter, and by a suitable choice of units the ratio could always be set to one, i.e. $(m_G/m_I) = 1$. Bessel (1784-1846) tested this ratio more accurately, and then in a definitive 1889 experiment E\"otv\"os was able to experimentally verify this equality of the inertial and gravitational masses to an accuracy of one part in $10^9$ \cite{Eotvos_1890,Eotvos_etal_1922,Bod_etal_1991}. 

Today, more than three hundred and twenty years after Newton proposed a comprehensive approach to studying the relation between the two masses of a body, this relation still continues to be the subject of modern theoretical and experimental investigations.  The question about the equality of inertial and passive gravitational masses arises in almost every theory of gravitation.  Nearly one hundred years ago, in 1915, the EP became a part of the foundation of Einstein's general theory of relativity; subsequently, many experimental efforts focused on testing the equivalence principle in the search for limits of general relativity.  Thus, the early tests of the EP were further improved by \cite{Roll_etal_1964} to one part in $10^{11}$. Most recently, a University of Washington group \cite{Baessler_etal_1999,Adelberger_2001} has improved upon Dicke's verification of the EP by several orders of magnitude, reporting $(m_G/m_I - 1) < 1.4 \times 10^{-13}$, thereby confirming Einstein's intuition.

\begin{figure}
  \vspace{-0pt}
  \begin{center}
    \includegraphics[width=0.50\textwidth]{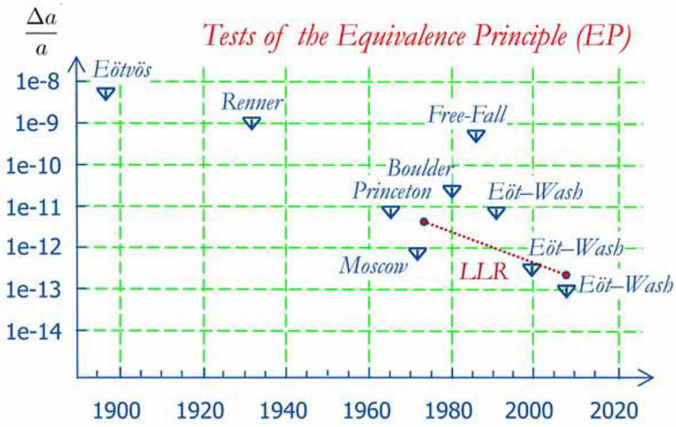}
  \end{center}
  \vspace{-5pt}
  \caption{The progress in the tests of the EP since the early 1900s  \cite{Turyshev-etal-2007}.
}
  \vspace{-10pt}
\end{figure}

Back in 1907, using the early version of the EP Einstein was already able to make important preliminary predictions regarding influence of gravity on light propagation; thereby, making the next important step in the development of his theory.  He realized that a ray of light coming from a distant star would appear to be attracted by solar mass while passing in the proximity of the Sun.  As a result, the ray's trajectory will be bent twice more in the direction towards the Sun compared to the same trajectory analyzed with the Newton's theory. In addition, light radiated by a star would interact with the star's gravitational potential, resulting in the radiation being slightly shifted toward the infrared end of the spectrum. 

About 1912, Einstein began a new phase of his gravitational research, with the help of his mathematician friend Marcel Grossmann, by phrasing his work in terms of the tensor calculus of Tullio Levi-Civita and Gregorio Ricci-Curbastro. The tensor calculus greatly facilitated calculations in four-dimensional space-time, a notion that Einstein had obtained from Hermann Minkowski's 1907 mathematical elaboration of Einstein's own special theory of relativity. Einstein called his new theory the general theory of relativity.  After a number of false starts, he published the definitive form of the field equations of his theory in late 1915 \cite{Einstein-1915,Einstein-1916}.  
Since that time, physicists have struggled to understand and verify various predictions of general theory of relativity with ever increasing accuracy.

\subsection{Test of Einstein's Equivalence Principle}
\label{sec:eep}

The Einstein's Equivalence Principle (EP)  \cite{Damour_Nordtvedt_93b,Williams-etal-2004,Williams-etal-2005} is at the foundation of general theory of relativity; therefore, testing the principle is very important. The EP includes three hypotheses: 
(i) 
universality of free fall (UFF), which states that freely falling bodies do have the same acceleration in the same gravitational field independent on their compositions (see Sec.~\ref{sec:eep}),
(ii) local Lorentz invariance (LLI), which suggests that clock rates are independent on the clock's velocities (see Sec.~\ref{LLI}), and
(iii) local position invariance (LPI), which postulates that clocks rates are also independent on their spacetime positions (see Sec.~\ref{LPI}). 
Using these three hypotheses Einstein deduced that gravity is a geometric property of spacetime \cite{Will-lrr-2006-3}. One can test both the validity of the EP and of the field equations that determine the geometric structure created by a mass distribution. There are two different ``flavors'' of the Principle,  the weak and the strong forms of the EP that are currently tested in various experiments performed with laboratory test masses and with bodies of astronomical sizes \cite{Williams-etal-2005}.

\subsubsection{The Weak  Equivalence Principle (WEP)}
\label{sec:wep}

The \emph{weak form of the EP} (the WEP) states that the gravitational properties of strong and electro-weak interactions obey the EP. In this case the relevant test-body differences are their fractional nuclear-binding differences, their neutron-to-proton ratios, their atomic charges, etc.  Furthermore, the equality of gravitational and inertial masses implies that different neutral massive test bodies will have the same free fall acceleration in an external gravitational field, and therefore in freely falling inertial frames the external gravitational field appears only in the form of a tidal interaction \cite{Singe-1960}. Apart from these tidal corrections, freely falling bodies behave as if external gravity was absent \cite{Anderson-etal-1996}. 

General relativity and other metric theories of gravity assume that the WEP is exact.  However, many extensions of the Standard Model that contain new macroscopic-range quantum fields predict quantum exchange forces that generically violate the WEP because they couple to generalized ``charges'' rather than to mass/energy as does gravity \cite{Damour_Nordtvedt_93b,Damour_Polyakov_94b,Damour_Piazza_Veneziano_02b}. 

In a laboratory, precise tests of the EP can be made by comparing the free fall accelerations, $a_1$ and $a_2$, of different test bodies. When the bodies are at the same distance from the source of the gravity, the expression for the equivalence principle takes the elegant form
\begin{equation} 
{\Delta a \over a} = {2(a_1- a_2) \over a_1 + a_2} =
\left[{m_G \over m_I}\right]_1 -\left[{m_G \over m_I}\right]_2 = \Delta\left[{m_G \over m_I}\right], 
\label{WEP_da} \end{equation}
\noindent where $m_G$ and $m_I$ are the gravitational and inertial masses of each body.

Currently, the most accurate results in testing the WEP were reported by ground-based laboratories \cite{Williams-etal-2005,Baessler_etal_1999}. The most recent result \cite{Adelberger_2001,Schlamminger-etal-2007} for the fractional differential acceleration between beryllium and titanium test bodies was given as
$\Delta a/a=(1.0 \pm 1.4) \times 10^{-13}$. The accuracy of these experiments is sufficiently high to confirm that the strong, weak, and electromagnetic interactions each contribute equally to the passive gravitational and inertial masses of the laboratory bodies. A review of the most recent laboratory tests of gravity can be found in Ref. \cite{Gundlach-etal-2007}. Significant improvements in the tests of the EP are expected from dedicated space-based experiments.

\begin{wrapfigure}{R}{0.19\textwidth}
  \vspace{-25pt}
  \begin{center}
    \includegraphics[width=0.19\textwidth]{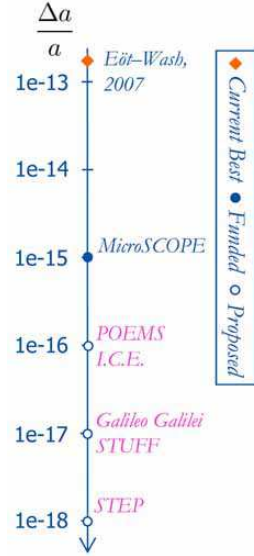}
  \end{center}
  \vspace{-10pt}
  \caption{Anticipated progress in the tests of the WEP \cite{Turyshev-etal-2007}.}
  \vspace{-15pt}
\label{fig:future-WEP-tests}
\end{wrapfigure}

The composition-independence of acceleration rates of various masses toward the Earth can be tested to many additional orders of magnitude precision in space-based laboratories, down to levels where some models of the unified theory of quantum  gravity, matter, and energy suggest a possible violation of the EP \cite{Damour_Nordtvedt_93b,Damour_Polyakov_94b,Damour_Piazza_Veneziano_02b}.  Note, in some scalar-tensor theories, the strength of EP violations and the magnitude of the fifth force mediated by the scalar can be drastically larger in space compared with that on the ground \cite{Mota-Barrow-2004-2}, which further justifies a space deployment.  Importantly, many of these theories predict observable violations of the EP at various levels of accuracy ranging from $10^{-13}$ down to $10^{-16}$. Therefore, even a confirmation of no EP-violation will be exceptionally valuable, placing useful constraints on the range of possibilities in the development of a unified physical theory. 

Compared with Earth-based laboratories, experiments in space can benefit from a range of conditions including free-fall and significantly reduced contributions due to seismic, thermal and many other sources of non-gravitational noise \cite{Turyshev-etal-2007}. As a result, there are many experiments proposed to test the EP in space. Below we present only a partial list of these missions.  Furthermore, to illustrate the use of different technologies, we present only the most representative concepts.  

The MicroSCOPE mission\footnote{\label{foot:microscope}Micro-Satellite \`a tra\^in\'ee Compens\'ee pour l'Observation du Principe d'Equivalence (MicroSCOPE), for more details, please see: \tt http://microscope.onera.fr/}  is a room-temperature EP experiment in space relying on electrostatic differential accelerometers \cite{Touboul-Rodrigues-2001}.  The mission is currently under development by CNES\footnote{Centre National d'Etudes Spatiales (CNES) -- the French Space Agency, see website at: {\tt http://www.cnes.fr/}} and ESA, scheduled for launch in 2010.  The design goal is to achieve a differential acceleration accuracy of $10^{-15}$. 

The Principle of Equivalence Measurement (PO\-EM) experiment
\cite{Reasenberg-Phillips-2007} is a ground-based test of the WEP, now under development.  It will be able to detect a violation of the EP with a fractional acceleration accuracy of 5 parts in 10$^{14}$ in a short (few days) experiment and 3 to 10 fold better in a longer experiment.  The experiment makes use of optical distance measurement (by TFG laser gauge \cite{Reasenberg-Phillips-2007}) and  will be  advantageously sensitive to short-range forces with a characteristic length scale of $\lambda < 10$~km. SR-POEM, a POEM-based  proposed room-temperature test of the WEP during a sub-orbital flight on a sounding rocket, was recently also presented \cite{Turyshev-etal-2007}. It is anticipated to be able to search for a violation of the EP with a single-flight  accuracy of one part in 10$^{16}$. Extension to higher accuracy in an orbital mission is under study. 

The Satellite Test of Equivalence Principle (STEP) mission \cite{step-2001-1,step-2001-2} is a proposed test of the EP to be conducted from a free-falling platform in space provided by a drag-free spacecraft orbiting the Earth.  STEP will test the composition independence of gravitational acceleration for cryogenically controlled test masses by searching for a violation of the EP with a fractional acceleration accuracy of one part in 10$^{18}$.  As such, this ambitious experiment will be able to test very precisely for the presence of any new non-metric, long range physical interactions.

\subsubsection{The Strong Equivalence Principle (SEP)}
\label{sec:sep}

In its \emph{strong form the EP} (the SEP) is extended to cover the gravitational properties resulting from gravitational energy itself \cite{Williams-etal-2005}.  In other words, it is an assumption about the way that gravity begets gravity, i.e. about the non-linear property of gravitation. Although general relativity assumes that the SEP is exact, alternate metric theories of gravity such as those involving scalar fields, and other extensions of gravity theory, typically violate the SEP. For the SEP case, the relevant test body differences are the fractional contributions to their masses by gravitational self-energy. Because of the extreme weakness of gravity, SEP test bodies must have astronomical sizes. 

Nordtvedt \cite{Ken_LLR68} suggested several solar system experiments for testing the SEP. One of these was the lunar test. Another, a search for the SEP effect in the motion of the Trojan asteroids, was carried out by
\cite{Orellana_Vucetich_1993}. Interplanetary spacecraft tests have been considered by \cite{Anderson-etal-1996} and discussed \cite{AndersonWilliams01}. An experiment employing existing binary pulsar data has been proposed \cite{Damour_Schafer_1991}. It was pointed out that binary pulsars may provide an excellent possibility for testing the SEP in the new regime of strong self-gravity \cite{Damour-EspFarese-1996-2}, however the corresponding tests have yet to reach competitive accuracy \cite{Kramer-etal-2006}.

The PPN formalism \cite{Will_book93} describes the motion of celestial bodies in a theoretical framework common to a wide class of metric theories of gravity. 
To facilitate investigation of a possible violation of the SEP, Eq.~(\ref{eq:4-26-mod}) allows for a possible inequality of the gravitational and inertial masses, given by the parameter $[{m_G}/{m_I}]_i$, which in the PPN formalism is expressed \cite{Nordtvedt_1968b} as
{}
\begin{equation}
\left[\frac{m_G}{m_I}\right]_{\tt SEP} = 1 + 
\eta\Big(\frac{E}{mc^2}\Big), \label{eq:MgMi} 
\end{equation}

\noindent where $m$ is the mass of a body, $E$ is the body's (negative) gravitational self-energy, $mc^2$ is its total mass-energy, and $\eta$ is a dimensionless constant for SEP violation \cite{Nordtvedt_1968b,Ken_LLR68}.
Any SEP violation is quantified by the parameter $\eta$: in fully-conservative, Lorentz-invariant theories of gravity \cite{Will_book93,Will-lrr-2006-3} the SEP parameter is related to the PPN parameters by $\eta = 4\beta - \gamma -3\equiv 4\bar\beta - \bar\gamma$. In general relativity $\gamma = \beta = 1$, so that $\eta = 0$ (see \cite{Williams-etal-2005,Will_book93,Will-lrr-2006-3}).

The quantity $E$ is the body's gravitational self-energy $(E< 0)$, which for a body $i$ is given by 
\begin{equation}
\left[{E  \over Mc^2}\right]_i= - \frac{G}{2m_ic^2}\int_i d^3 x d^3 x' 
\frac{\rho_i({\bf r})\rho_i({\bf r}')}{| {\bf r} - {\bf r}'|}.        
\label{eq:omega}
\end{equation}
For a sphere with a radius $R$ and uniform density, $E /mc^2 = -3Gm/5Rc^2 = -0.3 v_E^2/c^2$, where $v_E$ is the escape velocity. Accurate evaluation for solar system bodies requires numerical integration of the expression of Eq.~(\ref{eq:omega}). Evaluating the standard solar model \cite{Ulrich_1982} results in $(E /Mc^2)_\odot \sim -3.52 \times 10^{-6}$. Because gravitational self-energy is proportional to $m^2_i$ and also because of the extreme weakness of gravity, the typical values for the ratio $(E /mc^2)$ are $\sim 10^{-25}$ for bodies of laboratory sizes. Therefore, the experimental accuracy of a part in $10^{13}$ \cite{Adelberger_2001} which is so useful for the WEP is not sufficient to test on how gravitational self-energy contributes to the inertial and gravitational masses of small bodies. To test the SEP one must consider planetary-sized extended bodies, where the ratio Eq.~(\ref{eq:omega}) is considerably higher.

\begin{wrapfigure}{R}{0.19\textwidth}
  \vspace{-36pt}
  \begin{center}
    \includegraphics[width=0.19\textwidth]{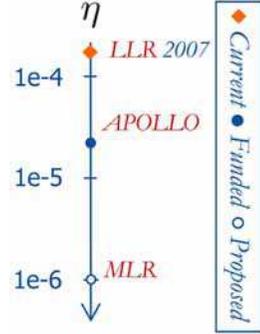}
  \end{center}
  \vspace{-00pt}
  \caption{Anticipated progress in the tests of the SEP \cite{Turyshev-etal-2007}.}
  \vspace{-18pt}
\end{wrapfigure}

Currently, the Earth-Moon-Sun system provides the best solar system arena for testing the SEP. LLR experiments involve reflecting laser beams off retroreflector arrays placed on the Moon by the Apollo astronauts and by an unmanned Soviet lander \cite{Williams-etal-2004,Williams-etal-2005}. Recent solutions using LLR data give $(-0.8 \pm 1.3) \times 10^{-13}$ for any possible inequality in the ratios of the gravitational and inertial masses for the Earth and Moon.  This result, in combination with laboratory experiments on the WEP, yields a SEP test of $(-1.8 \pm 1.9) \times 10^{-13}$  that corresponds to the value of the SEP violation parameter of $\eta =(4.0 \pm 4.3) \times 10^{-4}$. 
In addition, using the recent Cassini result for the PPN parameter $\gamma$, PPN parameter $\beta$ is determined at the level of $\bar\beta=(1.2\pm1.1)\times 10^{-4}$ (see details in \cite{Williams-etal-2004}).

With the new APOLLO\footnote{\label{ref:apollo}The Apache Point Observatory Lunar Laser-ranging Operations (APOLLO) is the new LLR station that was recently  built in New Mexico and  initiated operations in 2006.} facility \cite{Murphy-etal-2007,Williams-Turyshev-Murphy-2004}, the LLR science is going through a renaissance.  APOLLO's one-milli\-meter range precision will translate into order-of-mag\-nitude accuracy improvements in the test of the WEP and SEP (leading to accuracy at the level of ${\Delta a}/{a}\lesssim 1\times10^{-14}$ and $\eta\lesssim 2\times10^{-5}$ correspondingly), in the search for variability of Newton's gravitational constant (see Sec.~\ref{sec:variation-G}), and in the test of the gravitational inverse-square law (see Sec.~\ref{sec:inv-sq-law}) on scales of the Earth-moon distance (anticipated accuracy is $3\times10^{-11}$) \cite{Williams-Turyshev-Murphy-2004}.

The next step in this direction is interplanetary laser ranging \cite{laser-transponders-2006-1,laser-transponders-2006-2,Turyshev-Williams-2007}, for example, to a lander on Mars. Technology is available to conduct such measurements with a few picoseconds timing precision which could translate into mm-class accuracies achieved in ranging between the Earth and Mars. The resulting Mars Laser Ranging (MLR) experiment could test the weak and strong forms of the EP with accuracy at the $3\times10^{-15}$ and $2\times10^{-6}$ levels correspondingly, to measure the PPN parameter $\gamma$ (see Sec.~\ref{sec:mod-grav}) with accuracy below the $10^{-6}$ level, and to test gravitational inverse-square law at $\sim2$~AU distances with accuracy of $1\times 10^{-14}$, thereby greatly improving the accuracy of the current tests \cite{Turyshev-Williams-2007}. MLR could also advance research in several areas of science including remote-sensing geodesic and geophysical studies of Mars.

Furthermore, with the recently demonstrated capabilities of reliable laser links over large distances (e.g., tens of millions kilometers) in space \cite{laser-transponders-2006-1,laser-transponders-2006-2}, there is a strong possibility to improve the accuracy of gravity experiments with precision laser ranging over interplanetary scales \cite{Turyshev-Williams-2007}. Science justification for such an experiment is strong, the required technology is space-qualified and some components have already flown in space. With MLR, our very best laboratory for gravitational physics will be expanded to interplanetary distances, representing an upgrade in both scale and precision of this promising technique.

The experiments above are examples of the rich opportunities offered by the fundamental physics community to explore the validity of the EP. These experiments could potentially offer up to 5 orders of magnitude improvement over the  accuracy of the current tests of the EP.  Such experiments would dramatically enhance the range of validity for one of the most important physical principles or they could lead to a spectacular discovery.

\subsection{Testing Local Lorentz Invariance (LLI): Search for Physics Beyond the Standard Model}
\label{LLI}

The Standard Model coupled to general relativity is thought to be the effective low-energy limit of an underlying fundamental theory that unifies gravity and gravity and particle physics at the Planck scale. This underlying theory may well include Lorentz violation \cite{Colladay:1998fq} which could be detectable in space-based experiments \cite{Kostelecky:1994rn}.  Lorentz symmetry breaking due to non-trivial solutions of string field theory was first discussed in Ref. \cite{Kostelecky-Samuel:1989b}. These arise from the string field theory of open strings and may have implications for low-energy physics. For instance, assuming that the contribution of Lorentz-violating interactions to the vacuum energy is about half of the critical density implies that feeble tensor-mediated interactions in the range of $ \sim 10^{-4}$~m should exist \cite{Bertolami:1997,Bertolami-Paramos-Turyshev-2007} (see discussion in \cite{Turyshev-etal-2007}). 

\begin{figure}[h!]
  \vspace{-10pt}
  \begin{center}
    \includegraphics[width=0.5\textwidth]{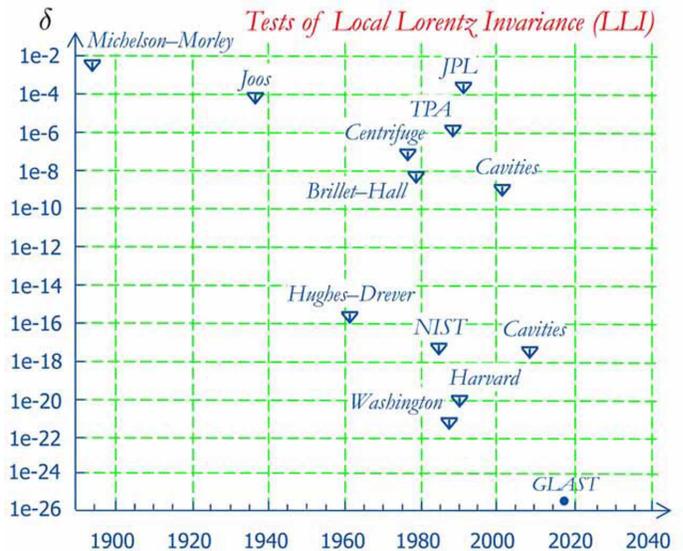}
  \end{center}
  \vspace{-10pt}
  \caption{The progress in the tests of the LLI since 1900s  \cite{Turyshev-etal-2007} and anticipated performance of the NASA's GLAST mission. 
}
  \vspace{-10pt}
\end{figure}

Limits on the violation of Lorentz symmetry are available from laser interferometric versions of the Michelson-Morley experiment, by comparing the velocity of light, $c$ and the maximum attainable velocity of massive particles, $c_i$, up to $\delta \equiv
|c^2/c_{i}^2 - 1| < 10^{-9}$ \cite{Brillet-Hall-1979}. More accurate tests can be performed via the Hughes-Drever experiment \cite{Hughes-etal:1960,Drever-1961}, where one searches for a time dependence of the quadrupole splitting of nuclear Zeeman levels along Earth's orbit. This technique achieves an impressive limit of $\delta < 3 \times 10^{-22}$ \cite{Lamoreaux-etal:1986}. A recent reassessment of these results reveals that more stringent bounds can be reached, up to 8 orders of magnitude higher \cite{Kostelecky:1999mr}. The parameterized post-Newtonian parameter $\alpha_{3}$ can be used to set astrophysical limits on the violation of momentum conservation and the existence of a preferred reference frame. This parameter, which vanishes in general relativity can be accurately determined from the pulse period of pulsars and millisecond pulsars \cite{Will-lrr-2006-3}. The most recent results yields a limit on the PPN parameter $\alpha_3$  of $|\alpha_{3}| < 2.2 \times 10^{-20}$ \cite{Bell:1995ax}. The NASA's GLAST misison is expected to improve the restuls of the LLI tests (see Fig.~\ref{fig:LPI-tests}) 

If one takes the Standard Model and adds appropriate terms that involve operators for Lorentz invariance violation \cite{Stecker:2001vb}, the result is the Standard-Model Extension (SME), which has provided a phenomenological framework for testing Lorentz-invariance
\cite{Kostelecky:2000hz,Kostelecky:1999mu}, and also suggested a number of new tests of relativistic gravity in the solar system \cite{Bailey:2006fd}. Compared with their ground-based analogs, space-based experiments in this area can provide improvements by as much as six orders of magnitude.  Several general reviews of the SME and corresponding efforts are available.   Recent studies of the ``aether theories'' \cite{Jacobson:2000xp} have shown that these models are naturally compatible with general relativity \cite{Will-lrr-2006-3}, but predict several non-vanishing Lorentz-violation parameters that could be measured in experiment.

Searches for extensions of special relativity on a space-based platforms are known as ``clock-comparison'' tests. The basic idea is to operate two or more high-precision clocks simultaneously and to compare their rates correlated with orbit parameters such as velocity relative to the microwave background and position in a gravitational environment.  The SME allows for the possibility that comparisons of the signals from different clocks will yield very small differences that can be detected in experiment. Currently, an experiment, called Atomic Clock Ensemble in Space (ACES), is aiming to do important tests of SME.  ACES is a European mission \cite{Cacciapuoti-etal:2007} in fundamental physics that will operate atomic clocks in the microgravity environment of the ISS with fractional frequency stability and accuracy of a few parts in 10$^{16}$.  ACES is is being prepared for a flight to the ISS in 2013-14 for the planned mission duration of 18 months \cite{Turyshev-etal-2007}.

\begin{figure}[h!]
  \vspace{-0pt}
  \begin{center}
    \includegraphics[width=0.44\textwidth]{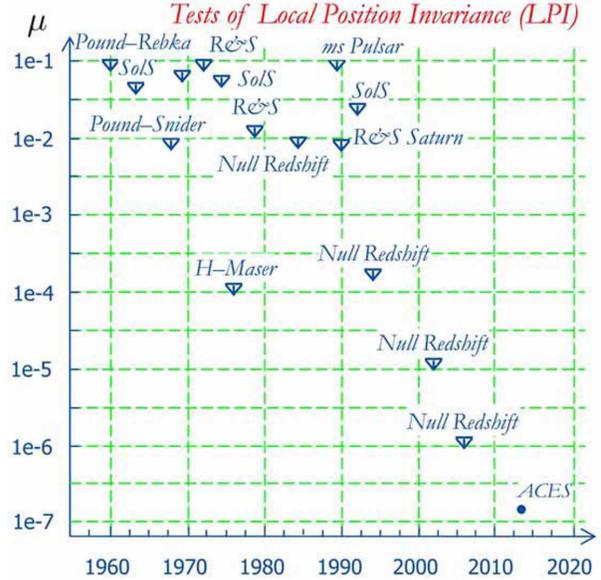}
  \end{center}
  \vspace{-10pt}
  \caption{The progress in the tests of the LPI since the early 1950s  \cite{Turyshev-etal-2007} and anticipated performance of the ESA's ACES mission. Here `SolS', for the tests with solar spectra, `R\&S' is that using rockets and spacecraft, and `Null Redshift' is for comparison of different atomic clocks (see \cite{Will-lrr-2006-3}).
}
  \vspace{-10pt}
\label{fig:LPI-tests}
\end{figure}

Optical clocks offer improved possibility of testing the time variations of fundamental constants at a high accuracy level (see \cite{Turyshev-etal-2007} and references therein).  Such measurements interestingly complement the tests of the LLI \cite{Wolf-Petit:1997}  and of the UFF to experimentally establish the validity of the EP. The universality of the gravitational red-shift can be tested at the same accuracy level by two optical clocks in free flight in a varying gravitational potential.  Constancy and isotropy of the speed of light can be tested by continuously comparing a space clock with a ground clock.  Optical clocks orbiting the Earth combined with a sufficiently accurate time and frequency transfer link, can improve present results by more than three orders of magnitude.

\subsection{Local Position Invariance (LPI)}
\label{LPI}

Given that both the WEP and LLI postulates have been tested with great accuracy, experiments concerning the universality of the gravitational red-shift measure the level to which the LPI holds. Therefore, violations of the LPI would imply that the rate of a free falling clock would be different when compared with a standard one, for instance on the Earth's surface. The accuracy to which the LPI holds as an invariance of Nature can be parameterized through $ \Delta \nu / \nu = (1 + \mu) U / c^2$.  The Pound-Rebka experiment, performed in 1960, further verified effects of gravity on light by testing the universality of gravity-induced frequency shift, $\Delta \nu$, that follows from the WEP:
${\Delta \nu / \nu} = {g h / c^2} = (2.57 \pm 0.26) \times 10^{-15},$ where $g$ is the acceleration of gravity and $h$ the height of fall \cite{Pound-Rebka-1960,Pound-Snider-1964}. This was the first convincing measurement of the gravitational red-shift made by Pound and Rebka in 1960. The experiment was based on M\"ossbauer Effect measurements between sources and detectors spanning the 22.5 m tower in the Jefferson Physical Laboratory at Harvard.  The Pound-Rebka experiment test LPI resulted in the limit of $\mu \simeq 10^{-2}$. In 1976, an accurate verification of the LPI was performed by Vessot and collaborators who compared the frequencies of two Hydrogen masers, one being on Earth and another one on a sub-orbital rocket.  The resulted Gravity Probe A experiment \cite{Vessot-etal-1980} exploited the much higher ``tower'' enabled by space; a sub-orbital Scout rocket carried a Hydrogen maser to an altitude of 10,273 km and a novel telemetry scheme allowed comparison with Hydrogen masers on the ground.  The clocks confirmed Einstein's prediction to 70 ppm, thereby establishing a limit of $ \vert \mu \vert < 2 \times 10^{-4}$. More than 30 years later, this remained the most precise measurement of the gravitational red-shift \cite{Will-lrr-2006-3}. Currently the the most stringent bound on possible violation of the LPI is $ \vert \mu \vert < 2.1 \times 10^{-5}$reported in \cite{Bauch-Weyers-2002}. The ESA's ACES misison is expected to improve the results of the LPI tests (see Fig.~\ref{fig:LPI-tests})

\subsection{Test of the variation of fundamental constants}
\label{sec:fest-fund-const}


Dirac's 70 year old idea of cosmic variation of physical constants has been revisited with the advent of models unifying the forces of nature based on the symmetry properties of possible extra dimensions, such as the Kaluza-Klein-inspired theories, Brans-Dicke theory, and supersymmetry models.  Alternative theories of gravity \cite{Will-lrr-2006-3} and theories of modified gravity \cite{Bertolami-Paramos-Turyshev-2007} include cosmologically evolving scalar fields that lead to variability of the fundamental constants. It has been  hypothesized that a variation of the cosmological scale factor with epoch could lead to temporal or spatial variation of the physical constants, specifically the gravitational constant, $G$, the fine-structure constant, $\alpha$, and the electron-proton mass ratio ($m_{\rm e}/m_{\rm p}$). 

In general, constraints on the variation of fundamental constants can be derived from a number of gravitational measurements, such as the test of the UFF, the motion of the planets in the solar system, stellar and galactic evolutions. They are based on the comparison of two time scales, the first (gravitational time) dictated by gravity (ephemeris, stellar ages, etc.), and the second (atomic time) determined by a non-gravitational system (e.g. atomic clocks, etc.) \cite{Canuto-Goldman-1982-2}. For instance, planetary and spacecraft ranging, neutron star binary observations, paleontological and primordial nucleosynthesis data allow one to constrain the relative variation of $G$ \cite{Uzan-2002}. Many of the corresponding experiments could reach a much higher precision if performed in space. 

\subsubsection{Fine-Structure Constant} 
\label{sec:variation-alpha}

The current limits on the evolution of $\alpha$ are established by laboratory measurements, studies of the abundances of radioactive isotopes and those of fluctuations in the cosmic microwave background, as well as other cosmological constraints (for review see \cite{Uzan-2002}).  Laboratory experiments are based on the comparison either of different atomic clocks or of atomic clocks with ultra-stable oscillators. They also have the advantage of being more reliable and reproducible, thus allowing better control of the systematics and better statistics compared with other methods. Their evident drawback is their short time scales, fixed by the fractional stability of the least precise standards. These time scales usually are of order of a month to a year so that the obtained constraints are restricted to the instantaneous variation today.  However, the shortness of the time scales is compensated by a much higher experimental sensitivity. 

There is a connection between the variation of the fundamental constants and a violation of the EP; in fact, the former almost always implies the latter.
For example, should there be an ultra-light scalar particle, its existence would lead to variability of the fundamental constants, such as $\alpha$ and $m_{\rm e}/m_{\rm p}$. Because  masses of nucleons are $\alpha$-dependent, by coupling to nucleons this particle would mediate an isotope-dependent long-range force \cite{Damour_Polyakov_94b,Dvali-Zaldarriaga-2002,Uzan-2002,Dent-2007}. The strength of the coupling is within a few of orders of magnitude from the existing experimental bounds for such forces; thus, the new force can be potentially measured in precision tests of the EP. Therefore, the existence of a new interaction mediated by a massless (or very low-mass) time-varying scalar field would lead to both the variation of the fundamental constants and violation of the WEP, ultimately resulting in observable deviations from general relativity.

Following the arguments above, for macroscopic bodies, one expects that their masses depend on all the coupling constants of the four known fundamental interactions, which has profound consequences concerning the motion of a body. In particular, because the $\alpha$-dependence is {\it a priori} composition-dependent, any variation of the fundamental constants will entail a violation of the UFF \cite{Uzan-2002}. This allows one to compare the ability of two classes of experiments -- clock-based and EP-testing ones -- to search for variation of the parameter $\alpha$ in a model-indepen\-dent way \cite{Nordtvedt-2002}. EP experiments have been superior performers. Thus, analysis of the frequency ratio of the 282-nm $^{199}$Hg$^+$ optical clock transition to the ground state hyperfine splitting in $^{133}$Cs was recently used to place a limit on its fractional variation of $\dot\alpha/\alpha \leq 1.3\times 10^{-16}~{\rm yr}^{-1}$ \cite{Fortier-etal-2007}. At the same time, the current accuracy of the EP tests \cite{Williams-etal-2005} already constrains the variation as $\Delta \alpha/\alpha \leq 10^{-10}\Delta U/c^2$, where $\Delta U$ is the change in the gravity potential. Therefore, for ground-based experiments (for which the variability in the gravitational potential is due to the orbital motion of the Earth) in one year the quantity $U_{\tt sun}/c^2$ varies by $1.66\times10^{-10}$, so a ground-based clock experiment must therefore be able to measure fractional frequency shifts between clocks to a precision of a part in $10^{20}$ in order to compete with EP experiments on the ground \cite{Nordtvedt-2002}. 

On the other hand, sending atomic clocks on a spacecraft to within a few solar radii of the Sun where the gravitational potential grows to $10^{-6} c^2$ could, however, be a competitive experiment if the relative frequencies of different on-board clocks could be measured to a precision better than a part in $10^{16}$. Such an experiment would allow for a direct measurement of any $\alpha$-variation, thus further motivating the development of  space-qualified clocks. With their accuracy surpassing the $10^{-17}$ level in the near future, optical clocks may be able to provide the needed capabilities  to directly test the variability of the fine-structure constant \cite{Turyshev-etal-2007}.

Clearly a solar fly-by on a highly-eccentric trajectory with very accurate clocks and inertial sensors makes for a compelling relativity test.  A potential use of highly-accurate optical clocks in such an experiment would likely lead to additional accuracy improvement in the tests of $\alpha$ and $m_{\rm e}/m_{\rm p}$, thereby providing a good justification for space deployment \cite{Schiller-etal-2006,Turyshev-etal-2007}. The resulting space-based laboratory experiment could lead to an important discovery.

\subsubsection{Gravitational Constant} 
\label{sec:variation-G}

A possible variation of Newton's gravitational constant $G$ could be related to the expansion of the universe depending on the cosmological model considered. Variability in $G$ can be tested in space with a much greater precision than on Earth \cite{Williams-etal-2004,Uzan-2002}.  
For example, a decreasing gravitational constant, $G$, coupled with angular momentum conservation is expected to increase a planet's semimajor axis, $a$, as $\dot a/a=-\dot G/G$.  The corresponding change in orbital phase grows quadratically with time, providing for strong sensitivity to the effect of $\dot G$.

\begin{wrapfigure}{R}{0.19\textwidth}
  \vspace{-30pt}
  \begin{center}
    \includegraphics[width=0.19\textwidth]{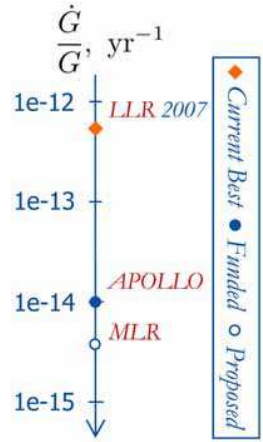}
  \end{center}
  \vspace{-10pt}
  \caption{Anticipated progress in the tests of possible variability in the gravitational constant \cite{Turyshev-etal-2007}.}
  \vspace{-10pt}
\label{fig:future-G-dot-tests}
\end{wrapfigure}

Space-based experiments using lunar and planetary ranging measurements currently are the best means to search for very small spatial or temporal gradients in the values of $G$ \cite{Williams-etal-2004,Williams-etal-2005}.  Thus, recent analysis of LLR data strongly limits such variations and constrains a local ($\sim$1~AU) scale expansion of the solar system as $\dot a/a=-\dot G/G =-(5\pm6) \times 10^{-13}$ yr$^{-1}$, including that due to cosmological effects \cite{Williams-etal-2007}. Interestingly, the achieved accuracy in $\dot G/G$ implies that, if this rate is representative of our cosmic history, then $G$ has changed by less than 1\% over the 13.4 Gyr age of the universe.

The ever-extending LLR data set and increase in the accuracy of lunar ranging (i.e., APOLLO) could lead to significant improvements in the search for variability of Newton's gravitational constant; an accuracy at the level of $\dot G/G\sim 1\times10^{-14}~{\rm yr}^{-1}$ is feasible with LLR \cite{Turyshev-Williams-2007}. High-accuracy timing measurements of binary and double pulsars could also provide a good test of the variability of the gravitational constant \cite{Nordtvedt-2002,Kramer-etal-2006}.  
Fig.~\ref{fig:future-G-dot-tests} shows anticipated progress in the tests of possible variability in the gravitational constant.

\subsection{\label{sec:inv-sq-law}Search for new physics via tests of the gravitational inverse square law}  

Many modern theories of gravity, including string, supersymmetry, and brane-world theories, have suggested that new physical interactions will appear at short ranges.  This may happen, in particular, because at sub-millimeter distances new dimensions can exist, thereby changing the gravitational inverse-square law \cite{ADD-1998} (for review of experiments, see \cite{Adelberger-Heckel-Nelson-2003}). Similar forces that act at short distances are predicted in supersymmetric theories with weak scale compactifications \cite{AnDD-1998}, in some theories with very low energy supersymmetry breaking \cite{Dimopoulos-Giudice-1998}, and also in theories of very low quantum gravity scale \cite{Sundrum-1999,Dvali-etal-1998}. These multiple predictions provide strong motivation for experiments that would test for possible deviations from Newton's gravitational inverse-square law at very short distances, notably on ranges from 1 mm to 1 $\mu$m.

An experimental confirmation of new fundamental for\-ces would provide an important insight into the physics beyond the Standard Model. A great interest in the subject was sparked after the 1986 claim of evidence for an intermediate range interaction with sub-gravitational strength \cite{Fischbach-Talmadge-1998}, leading to a wave of new experiments.

In its simplest versions, a new interaction (or a fifth force) would arise from the exchange of a light boson coupled to matter with a strength comparable to gravity. Planck-scale physics could give origin to such an interaction in a variety of ways,
thus yielding a Yukawa-type modification in the interaction energy between point-like masses. This new interaction can be derived, for instance, from extended supergravity theories after dimensional reduction \cite{Scherk}, compactification of
$5$-dimensional generalized Kaluza-Klein theories including gauge interactions at higher dimensions \cite{Bars-Visser-1986}, and also from string/M-theory. In general, the interaction energy, $V(r)$, between two point masses $m_1$ and $m_2$ can be expressed in terms
of the gravitational interaction as
{}
\begin{equation} 
V(r) = -  {G_{\infty}m_{1}m_{2} \over r}\big(1 +
\alpha\,e^{-r/\lambda}\big), \label{eq:2.1} 
\end{equation}
\noindent where $r = \vert {\bf r}_2 - {\bf r}_1 \vert$ is the distance between the masses, $G_{\infty}$ is the gravitational coupling for $r \rightarrow \infty$ and $\alpha$ and $\lambda$ are respectively the strength and range of the new interaction.
Naturally, $G_{\infty}$ has to be identified with Newton's gravitational constant and the gravitational coupling becomes dependent on $r$. Indeed, the force associated with Eq.~(\ref{eq:2.1}) is given by: ${\bf F}(r)={\vec \nabla} V(r)=  
 - G(r)m_{1}m_{2}\,\hat {\bf r}/r^2$,  where  
$G(r) = G_{\infty}\big[1 + \alpha\,(1 + r/\lambda)e^{-r/\lambda}\big]$.

Recent ground-based torsion-balance experiments \cite{Kapner:2006si} tested the gravitational inverse-square law at separations between 9.53 mm and $55~\mu$m, probing distances less than the dark-energy length scale $\lambda_d =\sqrt[4]{\hbar c/ u_d}\approx 85~\mu$m, with energy density $u_d\approx3.8~{\rm keV/cm}^3$.  It was found that the inverse-square law holds down to a length scale of $56~\mu$m and that an extra dimension must have a size less than $44~\mu$m (similar results were obtained by \cite{Tu-etal-2007}).  These results are important, as they signify the fact that modern experiments reached the level at which dark-energy physics can be tested in a laboratory setting; they also provided a new set of constraints on new forces \cite{Adelberger-etal-2007}, making such experiments very relevant and competitive with particle physics research. Also, recent laboratory experiments testing the Newton's second law for small accelerations \cite{Schlamminger-etal-2007,Gundlach-etal-2007} also provided useful constraints relevant to understanding several current astrophysical puzzles.  

Sensitive experiments searching for weak forces invariably require soft suspension for the measurement degree of freedom.  A promising soft suspension with low dissipation is superconducting magnetic levitation.  Levitation in 1-{\sl g}, however, requires a large magnetic field, which tends to couple to the measurement degree of freedom through metrology errors and coil non-linearity, and stiffen the mode.  The high magnetic field will also make suspension more dissipative.  The situation improves dramatically in space.  The {\sl g}-level is reduced by five to six orders of magnitude, so the test masses can be supported with weaker magnetic springs, permitting the realization of both the lowest resonance frequency and lowest dissipation.  The microgravity conditions also allow for an improved design of the null experiment, free from the geometric constraints of the torsion balance.

The Inverse-Square Law Experiment in Space (ISLES) is a proposed experiment whose objective is to perform a highly accurate test of Newton's gravitational law in space \cite{Paik-etal-2007}. ISLES combines the advantages of the microgravity environment  with superconducting accelerometer technology to improve the current ground-based limits in the strength of violation \cite{Chiaverini-etal-2003} by four to six orders of magnitude in the range below $100~\mu$m.  The experiment will be sensitive enough to probe large extra dimensions down to $5~\mu$m and also to probe the existence of the axion which, if it exists, is expected to violate the inverse-square law in the range accessible by ISLES.

The recent theoretical ideas concerning new particles and new dimensions have reshaped the way we think about the universe. Thus, should the next generation of experiments detects a force violating the inverse-square law, such a discovery would imply the existence of either an extra spatial dimension, or a massive graviton, or the presence of a new fundamental interaction \cite{Adelberger-etal-2007}.   

While most attention has focused on the behavior of gravity at short distances, it is possible that tiny deviations from the inverse-square law occur at much larger distances. In fact, there is a possibility that non-compact extra dimensions could produce such deviations at astronomical distances \cite{Dvali-Gruzinov-Zaldarriaga-2003} (for discussion see Sec.~\ref{sec:mod-grav}).

By far the most stringent constraints on a test of the inverse-square law to date come from very precise measurements of the Moon's orbit about the Earth. Even though the Moon's orbit has a mean radius of 384,000 km, the models agree with the data at the level of 4 mm! As a result, analysis of the LLR data tests the gravitational inverse-square law to $3\times10^{-11}$ of the gravitational field strength on scales of the Earth-moon distance \cite{Williams-Turyshev-Murphy-2004}.

Interplanetary laser ranging could provide conditions that are needed to improve the tests of the inverse-square law on the interplanetary scales \cite{Turyshev-Williams-2007}. MLR could be used to perform such an experiment that could reach the accuracy of $1\times 10^{-14}$ at 2~AU distances, thereby improving the current tests  by several orders of magnitude. 

Although most of the modern experiments do not show disagreements with Newton's law, there are puzzles that require further investigation. The radiometric tracking data received from the Pioneer 10 and 11 spacecraft at heliocentric distances between 20 and 70 AU has consistently indicated the presence of a small, anomalous, Doppler drift in the spacecraft carrier frequency. The drift can be interpreted as due to a constant sunward acceleration of $a_{\rm P}  = (8.74 \pm 1.33)\times 10^{-10}~{\rm m/s}^2$ for each particular craft \cite{pious}. This apparent violation of the inverse-square law has become known as the Pioneer anomaly \cite{Turyshev-etal-2007}.

The possibility that the anomalous behavior will continue to defy attempts at a conventional explanation has resulted in a growing discussion about the origin of the discovered effect. A recently initiated investigation of the anomalous signal using the entire record of the Pioneer spacecraft telemetry files in conjunction with the analysis of a much extended Pioneer Doppler data may soon reveal the origin of the  anomaly \cite{Turyshev-etal-2007,pio-data}. 

\subsection{\label{sec:mod-grav}
Tests of alternative and modified gravity theories with gravitational experiments in space}

Given the immense challenge posed by the unexpected discovery of the accelerated expansion of the universe, it is important to explore every option to explain and probe the underlying physics.  Theoretical efforts in this area offer a rich spectrum of new ideas, some of them are discussed below, that can be tested by experiment.

Motivated by the dark energy and dark matter problems, long-distance gravity modification is one of the radical proposals that has recently gained attention \cite{Deffayet-Dvali-Gabadadze-2002}.  Theories that modify gravity at cosmological distances exhibit a strong coupling phenomenon of extra graviton polarizations \cite{Deffayet-etal-2002,Dvali-2006}. This strong coupling phenomenon plays an important role for this class of theories in allowing them to agree with solar system constraints.  In particular, the ``brane-induced gravity'' model \cite{Dvali-Gabadadze-Porrati-2003} provides a new and interesting way of modifying gravity at large distances to produce an accelerated expansion of the universe, without the need for a non-vanishing cosmological constant \cite{Deffayet:2000uy,Deffayet-Dvali-Gabadadze-2002}. One of the peculiarities of this model is the way one recovers the usual gravitational interaction at small (i.e. non-cosmological) distances, motivating precision tests of gravity on solar system scales \cite{Bekenstein:2006fi}.   

The Eddington parameter $\gamma$, whose value in general relativity is unity, is perhaps the most fundamental PPN parameter \cite{Will_book93,Will-lrr-2006-3}, in that $\frac{1}{2}\bar\gamma$ is a measure, for example, of the fractional strength of the scalar gravity interaction in scalar-tensor theories of gravity \cite{Damour_Nordtvedt_93b}.  Currently, the most precise value for this parameter, $\bar\gamma = (2.1\pm2.3)\times10^{-5}$, was obtained using radio-metric tracking data received from the Cassini spacecraft \cite{Bertotti-Iess-Tortora-2003} during a solar conjunction experiment.\footnote{A similar experiment is planned for the ESA's {BepiColombo} mission to Mercury \cite{Iess-Asmar:2007}.} This accuracy approaches the region where multiple tensor-scalar gravity models, consistent with the recent cosmological observations \cite{Spergel-etal-2006}, predict a lower bound for the present value of this parameter at the level of $\bar\gamma \sim 10^{-6}-10^{-7}$ \cite{Damour-EspFarese-1996-2,Damour_Nordtvedt_93b,Damour_Polyakov_94b,Damour_Piazza_Veneziano_02b}.  Therefore, improving the measurement of this parameter would provide crucial information to separate modern scalar-tensor theories of gravity from general relativity, probe possible ways for gravity quantization, and test modern theories of cosmological evolution.

Interplanetary laser ranging could lead to a significant improvement in the accuracy of the parameter $\gamma$. Thus, precision ranging between the Earth and a lander on Mars during solar conjunctions may offer a suitable opportunity (i.e., MLR). If the lander were to be equipped with a laser transponder capable of reaching a precision of 1~cm, a measurement of $\gamma$ with accuracy of 1 part in 10$^6$ is possible. To reach accuracies beyond this level one must rely on a dedicated space experiment \cite{Turyshev-etal-2007,Turyshev-Williams-2007}.

\begin{wrapfigure}{R}{0.19\textwidth}
  \vspace{-30pt}
  \begin{center}
    \includegraphics[width=0.19\textwidth]{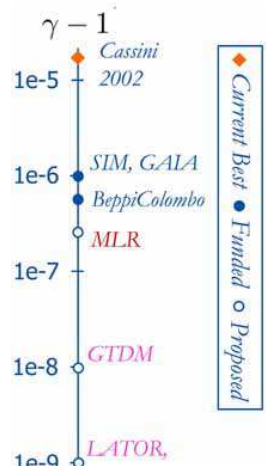}
  \end{center}
  \vspace{-20pt}
  \caption{Anticipated progress in the tests of the PPN parameter $\gamma$.}
  \vspace{-10pt}
\label{fig:future-gamma-tests}
\end{wrapfigure}

The Gravitational Time Delay Mission (GTDM) \cite{Ashby-Bender-2006} proposes to use laser ranging between two drag-free spacecraft (with spurious acceleration levels below $1.3 \times  10^{-13}~{\rm m/s}^2/\sqrt{\rm Hz}$ at  $0.4~\mu$Hz) to accurately measure the Shapiro time delay for laser beams passing near the Sun. One spacecraft would be kept at the L1 Lagrange point of the Earth-Sun system with the other one being placed on a 3:2 Earth-resonant, LATOR-type, orbit (see \cite{Turyshev-etal-2006} for details). A high-stability frequency standard ($\delta f/f\lesssim 1 \times  10^{-13}~1/\sqrt{\rm Hz}$ at $0.4~\mu$Hz) located on the L1 spacecraft permits accurate measurement of the time delay. If requirements on the performance of the disturbance compensation system, the timing transfer process, and high-accuracy orbit determination are successfully addressed, then determination of the time delay of interplanetary signals to 0.5 ps precision in terms of the instantaneous clock frequency could lead to an accuracy of 2 parts in $10^{8}$ in measuring the parameter $\gamma$.

The Laser Astrometric Test of Relativity (LATOR) \cite{Turyshev-etal-2006} proposes to measure the parameter $\gamma$ with accuracy of a part in 10$^9$, which is a factor of 30,000 beyond the currently best Cassini's 2003 result \cite{Bertotti-Iess-Tortora-2003}.  The key element of LATOR is a geometric redundancy provided by the long-baseline optical interferometry and interplanetary laser ranging. 
By using a combination of independent time-series of gravitational deflection of light in the immediate proximity to the Sun, along with measurements of the Shapiro time delay on interplanetary scales (to a precision better than 0.01 picoradians and 3 mm, respectively), LATOR will significantly improve our knowledge of relativistic gravity and cosmology. LATOR's primary measurement, precise observation of the non-Euclidean geometry of a light triangle that surrounds the Sun, pushes to unprecedented accuracy the search for cosmologically relevant scalar-tensor theories of gravity by looking for a remnant scalar field in today's solar system.  LATOR could lead to very robust advances in the tests of fundamental physics -- it could discover a violation or extension of general relativity or reveal the presence of an additional long range interaction.
 
\section{Conclusions}
\label{sec:conclusions}

The nature of gravity is fundamental to the understanding of the solar system and the large scale structure of the Universe. This importance motivates various precision tests of gravity both in laboratories and in space. It also stimulates technology developments that will be critical for new research.   Thus, recently, a new generation of high performance quantum sensors (ultra-stable atomic clocks, accelerometers, gyroscopes, gravimeters, gravity gradiometers, etc.) surpassed previous state-of-the-art instruments, demonstrating the high potential of these techniques based on the engineering and manipulation of atomic systems.  Atomic clocks and inertial quantum sensors represent a key technology for accurate frequency measurements and ultra-precise monitoring of accelerations and rotations.  New quantum devices based on ultra-cold atoms will enable unique fundamental physics experiments in the future. This progress promises very important new results in gravitational research for the next decade \cite{Turyshev-etal-2007}. 

\section*{Acknowledgments}

The work described here was carried out at the Jet Propulsion Laboratory, California Institute of Technology, under a contract with the National Aeronautics and Space Administration.


\begin{thebibliography}{100}

\bibitem{Einstein-1915}
A. {Einstein}, Sitzungsberichte der Preussischen Akademie der Wissenschaften zu
  Berlin  844  (1915).

\bibitem{Einstein-1916}
A. {Einstein}, Annalen der Physik {\bf 49},  146  (1916).

\bibitem{Turyshev-etal-2007}
S.~G. {Turyshev} {\it et~al.}, Int.J.Mod.Phys. D {\bf 16},  1879  (2007), arXiv:0711.0150 [gr-qc].

\bibitem{Dyson-Eddington-Davidson-1920}
F.~W. {Dyson}, A.~S. {Eddington}, and C. {Davidson}, Philos. Trans. Royal Soc.
  London {\bf 220A},  291  (1920).

\bibitem{Coles-2001}
P. {Coles},  in {\em Historical Development of Modern Cosmology}, Vol.~252 of
  {\em Astronomical Society of the Pacific Conference Series}, edited by V.~J.
  {Mart{\'{\i}}nez}, V. {Trimble}, and M.~J. {Pons-Border{\'{\i}}a}
  (Astronomical Society of the Pacific, San Francisco, 2001), pp.\ 21--41.

\bibitem{Kennefick-2007}
D.~{Kennefick}, arXiv:0709.0685 [physics.hist-ph].


\bibitem{viking_shapiro1}
I.~I. Shapiro, C.~C. Counselman, and R.~W. King, Phys. Rev. Lett. {\bf 36},   555  (1976).

\bibitem{viking_shapiro2}
I.~I. {Shapiro} {\it et~al.}, J. Geophys. Res. {\bf 82},  4329  (1977).


\bibitem{Anderson_etal_02}
J.~D. {Anderson} {\it et~al.}, BAAS {\bf 34},  833  (2002).

\bibitem{Pitjeva-2005}
E.~V. {Pitjeva}, Solar System Research {\bf 39},  176  (2005).


\bibitem{Lebach95}
D.~E. Lebach {\it et~al.}, Phys. Rev. Lett. {\bf 75},  1439  (1995).


\bibitem{Shapiro_SS_etal_2004}
S.~S. Shapiro, J.~L. Davis, D.~E. Lebach, and J.~S. Gregory, Phys. Rev. Lett.
  {\bf 92},  121101  (2004).

\bibitem{Williams-etal-2004}
J.~G. Williams, S.~G. Turyshev, and D.~H. Boggs, Phys. Rev. Lett. {\bf 93},
  261101  (2004), gr-qc/0411113.

\bibitem{Williams-etal-2005}
J.~G. {Williams}, S.~G. {Turyshev}, and D.~H. {Boggs},  
gr-qc/0507083.

\bibitem{Iess_etal_1999}
L. {Iess}, G. {Giampieri}, J.~D. {Anderson}, and B. {Bertotti}, Class. Quant.
  Grav. {\bf 16},  1487  (1999).

\bibitem{Bertotti-Iess-Tortora-2003}
B. Bertotti, L. Iess, P. Tortora, Nature {\bf 425},  374  (2003).

\bibitem{DamourTaylor92}
T. Damour, J.~H. Taylor, Phys. Rev. D {\bf 45},  1840  (1992).

\bibitem{Taylor_etal92}
J.~H. {Taylor}, A. {Wolszczan}, T. {Damour}, and J.~M. {Weisberg}, Nature {\bf
  355},  132  (1992).

\bibitem{Damour-EspFarese-1996-2}
T. Damour and G. Esposito-Farese, Phys. Rev. D {\bf 54},  1474  (1996);
%
T. Damour and G. Esposito-Farese, Phys. Rev. D {\bf 58},  042001  (1998).

\bibitem{Lange_etal_2001}
C. Lange {\it et~al.}, MNRAS {\bf 326},  274  (2001).

\bibitem{Will-lrr-2006-3}
C.~M. Will, Living Rev. Relativity {\bf 9},    (2006).

\bibitem{Damour-2006-pdg}
W.~M. Yao {\it et~al.}, J. Phys. G {\bf 33},  205  (2006).

\bibitem{Turyshev:1996}
S.~G. Turyshev, JPL Publication \#96-13. Pasadena, CA (1996), gr-qc/9606063.

\bibitem{Moyer-1981-2}
T.~D. {Moyer}, Celestial Mechanics {\bf 23},  57  (1981).

\bibitem{Standish_etal_92}
E.~M. Standish, X.~X. Newhall, J.~G. Williams, and D.~K. Yeomans,  in {\em
  Orbital Ephemerides of the Sun, Moon, and Planets}, {\em Explanatory
  Supplement to the Astronomical Almanac}, P.~K. Seidelmann, ed. (University Science Books, Mill Valley, 1992), pp.\
  279--323.

\bibitem{Will_book93}
C.~M. Will, {\em Theory and Experiment in Gravitational Physics} (Cambridge
  University Press, 1993).

\bibitem{Moyer-2003}
T.~D. {Moyer}, {\em Formulation for Observed and Computed Values of Deep Space
  Network Data Types for Navigation}, {\em JPL Deep-Space Communications and
  Navigation Series} (John Wiley \& Sons, Inc., Hoboken, New Jersey, 2003).

\bibitem{Turyshev_etal_acfc_2003}
S.~G. {Turyshev} {\it et~al.},  in {\em Astrophysics, Clocks and Fundamental
  Constants}, Vol.~648 of {\em Lect. Notes Phys.}, S.~G.
  {Karshenboim} and E. {Peik}, eds. (Springer Verlag, Berlin, 2004), pp.\ 311--330, gr-qc/0311039.

\bibitem{Brans-Dicke-1961}
C. Brans and R.~H. Dicke, Phys. Rev. {\bf 124},  925  (1961).

\bibitem{Damour_Nordtvedt_93b}
T. Damour, K. Nordtvedt, Phys. Rev. D {\bf 48},  3436  (1993).

\bibitem{Witten:2001}
E. {Witten},  in {\em APS/DPF/DPB Summer Study on the Future of Particle
  Physics}, edited by N. {Graf} (eConf C010630, paper \#337, Snowmass,
  Colorado, 2001).

\bibitem{Witten:2003}
E. {Witten},  in {\em The future of theoretical physics and cosmology}, edited
  by G.~W. Gibbons, E.~P.~S. Shellard, and S.~J. Rankin (Cambridge University
  Press, 2003), pp.\ 455--462.

\bibitem{Weinberg-1989:RevModPhys}
S. Weinberg, Rev. Mod. Phys. {\bf 61},  1  (1989).

\bibitem{Green-Schwarz-Witten-1987}
M. Green, J. Schwarz, and E. Witten, {\em Superstring Theory} (Cambridge
  University Press, 1987).

\bibitem{Polchinski-95}
J. Polchinski, Phys. Rev. Lett. {\bf 75},  4724  (1995).

\bibitem{Horava:1995qa}
P. Horava and E. Witten, Nucl. Phys. B {\bf 460},  506  (1996).

\bibitem{Lukas-1999}
A. Lukas, B.~A. Ovrut, and D. Waldram, Phys. Rev. D {\bf 60},  086001  (1999).

\bibitem{Fradkin-Tseytlin-1987}
E.~S. Fradkin and A.~A. Tseytlin, Nucl. Phys. B {\bf 201},  469  (1982).

\bibitem{Avramidi-Barvinsky-1982}
G. Avramidi and A.~O. Barvinsky, Phys. Lett. B {\bf 159},  269  (1985).

\bibitem{Goldman-etal-1992}
T. Goldman, J. P\'erez-Mercader, F. Cooper, and M.~M. Nieto, Phys. Lett. {\bf
  281},  219  (1992).

\bibitem{Damour_Polyakov_94b}
T. Damour and A.~M. Polyakov, Nucl. Phys. B {\bf 423},  532  (1994).

\bibitem{Damour_Piazza_Veneziano_02b}
T. Damour, F. Piazza, and G. Veneziano, Phys. Rev. D {\bf 66},  046007  (2002).

\bibitem{Turyshev-etal-2006}
S.~G. Turyshev, M. Shao, and K. Nordtvedt,  in {\em Lasers, Clocks, and
  Drag-Free: Technologies for Future Exploration in Space and Tests of
  Gravity}, edited by H. Dittus, C. Laemmerzahl, and S.~G. Turyshev (Springer
  Verlag, Berlin, 2006), pp.\ 429--493, gr-qc/0601035.

%
%

\bibitem{Turyshev-LATOR:2003}
S.~G. {Turyshev}, M. {Shao}, and K. {Nordtvedt}, Class. Quant. Grav. {\bf 21},
  2773  (2004), gr-qc/0311020.


\bibitem{Carroll:2004de}
S.~M. Carroll {\it et~al.}, Phys. Rev. D {\bf 71},  063513  (2005).

\bibitem{Navarro:2005da}
I. Navarro and K. Van~Acoleyen, JCAP {\bf 0603},  008  (2006).

\bibitem{Chiba:2003ir}
T. Chiba, Phys. Lett. {\bf B575},  1  (2003).

\bibitem{Dvali-Gabadadze-Porrati-2003}
G. Dvali, G. Gabadadze, and M. Porrati, Phys. Lett. B {\bf 485},  208  (2000).

\bibitem{Perlmutter:1998np}
S. {Perlmutter} {\it et~al.}, Astrophys. J. {\bf 517},  565  (1999).

\bibitem{Riess_supernovae98}
A.~G. Riess {\it et~al.}, Astron. J. {\bf 116},  1009  (1998).

\bibitem{Dvali-Gruzinov-Zaldarriaga-2003}
G. Dvali, A. Gruzinov, and M. Zaldarriaga, Phys. Rev. D {\bf 68},  024012
  (2003).

\bibitem{Cembranos:2005fi}
J.~A.~R. Cembranos, Phys. Rev. D {\bf 73},  064029  (2006).

\bibitem{Erickcek:2006vf}
A.~L. Erickcek, T.~L. Smith, and M. Kamionkowski, Phys. Rev. D {\bf 74},  121501
   (2006).

\bibitem{Nojiri:2006gh}
S. Nojiri and S.~D. Odintsov, Phys. Rev. D {\bf 74},  086005  (2006).

\bibitem{Wetterich:2004ss}
C. Wetterich, Phys. Rev. Lett. {\bf 90},  231302  (2003).

\bibitem{Peebles:2002gy}
P.~J.~E. Peebles, B. Ratra, Rev.Mod.Phys. {\bf 75},  559  (2003).

\bibitem{Wetterich:2004pv}
C. Wetterich, Phys. Lett. B {\bf 594},  17  (2004).

\bibitem{Ferreira_Joyce:1997}
P.~G. {Ferreira}, M. {Joyce}, Phys. Rev. Lett. {\bf 79},  4740  (1997).

\bibitem{Albrecht_Skordis:2000}
A. {Albrecht}, C. {Skordis}, Phys. Rev. Lett. {\bf 84},  2076
  (2000).

\bibitem{Bento_Bertolami_Santos:2002}
M.~C. {Bento}, O. {Bertolami}, and N.~C. {Santos}, Phys. Rev. D {\bf 65},
  067301  (2002).

\bibitem{Ferreira_Joyce:1998}
P.~G. {Ferreira}, M. {Joyce}, Phys. Rev. D {\bf 58},  023503  (1998).

\bibitem{Caldwell:1999ew}
R.~R. Caldwell, Phys. Lett. B {\bf 545},  23  (2002);
%
R.~R. Caldwell and E.~V. Linder, Phys. Rev. Lett. {\bf 95},  141301  (2005).

\bibitem{Cline:2003gs}
J.~M. Cline, S. Jeon, and G.~D. Moore, Phys. Rev. D {\bf 70},  043543  (2004).

\bibitem{Rubakov:2008nh}
V.~A. Rubakov, P.~G. Tinyakov, arXiv:0802.4379 [hep-th].

\bibitem{Sergienko:2008tf}
S. Sergienko, V. Rubakov, arXiv:0803.3163 [hep-th].

\bibitem{Sreekumar_etal:1998}
P. {Sreekumar} {\it et~al.}, Astrophys. J. {\bf 494},  523  (1998).

\bibitem{ArkaniHamed:2003uz}
N. Arkani-Hamed, P. Creminelli, S. Mukohyama, and M. Zaldarriaga, JCAP {\bf
  0404},  001  (2004).

\bibitem{Piazza:2004df}
F. Piazza and S. Tsujikawa, JCAP {\bf 0407},  004  (2004).

\bibitem{Eotvos_1890}
R.~v. {E\"otv\"os}, Mathematische und Naturwissenschaftliche Berichte aus
  Ungarn {\bf 8},  65  (1890).

\bibitem{Eotvos_etal_1922}
R.~v. {E\"otv\"os}, D. {Pek\'ar}, and E. {Fekete}, Ann. Phys. (Leipzig) {\bf
  68},  11  (1922), {.English translation for the U. S. Department of Energy by
  J. Achzenter, M. {Bickeb\"oller}, K. {Br\"auer}, P. Buck, E. Fischbach, G.
  Lubeck, C. Talmadge, University of Washington preprint 40048-13-N6. - More
  complete English text reprinted earlier in Annales Universitatis Scientiarium
  Budapestiensis de Rolando {E\"otv\"os} Nominate, Sectio Geologica {\bf 7},
  111 (1963)}.

\bibitem{Bod_etal_1991}
L. Bod, E. Fischbach, G. Marx, and M. N\'aray-Ziegler, Acta Physica Hungarica
  {\bf 69},  335  (1991).

\bibitem{Roll_etal_1964}
P.~G. {Roll}, R. {Krotkov}, and R.~H. {Dicke}, Ann. Phys. (N.Y.) {\bf 26},  442
   (1964).

\bibitem{Baessler_etal_1999}
S. Bae\ss{}ler {\it et~al.}, Phys. Rev. Lett. {\bf 83},  3585  (1999).

\bibitem{Adelberger_2001}
E. Adelberger, Class. Quantum Grav. {\bf 18},  2397  (2001).

\bibitem{Singe-1960}
J.~L. {Singe}, {\em Relativity: the General Theory} (Amsterdam: North-Holland, 1960).

\bibitem{Anderson-etal-1996}
J.~D. Anderson, M. Gross, K.~L. Nordtvedt, and S.~G. Turyshev, Astrophys. J.
  {\bf 459},  365  (1996).

\bibitem{Schlamminger-etal-2007}
S. Schlamminger {\it et~al.}, Phys. Rev. Lett. {\bf 100},  041101  (2008).

\bibitem{Gundlach-etal-2007}
J.~H. {Gundlach} {\it et~al.}, Phys. Rev. Lett. {\bf 98},  150801  (2007).

\bibitem{Mota-Barrow-2004-2}
D.~F. Mota, J.~D. Barrow, MNRAS {\bf 349},  291
  (2004);
%
J. Khoury, A. Weltman, Phys. Rev. D {\bf 69},  044026  (2004);
%
D.~F. Mota, D.~J. Shaw, Phys. Rev. D {\bf 75},  063501  (2007).

\bibitem{Touboul-Rodrigues-2001}
P. {Touboul} and M. {Rodrigues}, Class. Quant. Grav. {\bf 18},  2487  (2001).

\bibitem{Reasenberg-Phillips-2007}
R.~D. Reasenberg and J.~D. Phillips, Int. J. Mod. Phys. D {\bf 16},  2245 (2007);
%
J.~D. Phillips and R.~D. Reasenberg, Rev. Sci. Instr. {\bf 76},  064501 (2005).

\bibitem{step-2001-1}
J. {Mester} {\it et~al.}, Class. Quant. Grav. {\bf 18},  2475  (2001).

\bibitem{step-2001-2}
P. {Worden}, J. {Mester}, and R. {Torii}, Class. Quant. Grav. {\bf 18},  2543
  (2001).

\bibitem{Ken_LLR68}
K. {Nordtvedt}, Jr., Phys. Rev. {\bf 170},  1186  (1968).

\bibitem{Orellana_Vucetich_1993}
R.~B. {Orellana} and H. {Vucetich}, Astron. Astrophys. {\bf 273},  313  (1993).

\bibitem{AndersonWilliams01}
J.~D. {Anderson} and J.~G. {Williams}, Class. Quant. Grav. {\bf 18},  2447
  (2001).

\bibitem{Damour_Schafer_1991}
T. Damour, G. Sch\"afer, Phys. Rev. Lett. {\bf 66},  2549  (1991).

\bibitem{Kramer-etal-2006}
M. Kramer {\it et~al.}, Science {\bf 314},  97  (2006).

\bibitem{Nordtvedt_1968b}
K. Nordtvedt, Phys. Rev. {\bf 169},  1017  (1968).

\bibitem{Ulrich_1982}
R.~K. Ulrich, Astrophys. J. {\bf 258},  404  (1982).

\bibitem{Murphy-etal-2007}
T.~W. {Murphy} {\it et~al.}, Int.J.Mod.Phys. D {\bf 16},  2127  (2007).

\bibitem{Williams-Turyshev-Murphy-2004}
J.~G. {Williams}, S.~G. {Turyshev}, and T.~W. {Murphy}, Int. J. Mod. Phys. D {\bf 13},  567  (2004).

\bibitem{laser-transponders-2006-1}
D.~E. Smith {\it et~al.}, Science {\bf 311},  53  (2006).

\bibitem{laser-transponders-2006-2}
X. Sun {\it et~al.},  in {\em OSA Annual Meeting Abstracts, Tucson, AZ, Oct.
  16-20, 2005} (OSA, 2005).

\bibitem{Turyshev-Williams-2007}
S.~G. Turyshev and J.~G. Williams, Int. J. Mod. Phys. D {\bf 16},  2165 (2007).

\bibitem{Colladay:1998fq}
D. Colladay and V.~A. Kostelecky, Phys. Rev. D {\bf 58},  116002  (1998);
%
S.~R. Coleman and S.~L. Glashow, Phys. Rev. D {\bf 59},  116008  (1999).

\bibitem{Kostelecky:1994rn}
V.~A. {Kosteleck{\'y}} and R. {Potting}, Phys. Rev. D {\bf 51},  3923  (1995).

\bibitem{Kostelecky-Samuel:1989b}
V.~A. {Kosteleck\'y} and S. {Samuel}, Phys. Rev. Lett. {\bf 63},  224  (1989).

\bibitem{Bertolami:1997}
O. {Bertolami}, Class. Quant. Grav. {\bf 14},  2785  (1997).

\bibitem{Bertolami-Paramos-Turyshev-2007}
O. Bertolami, J. Paramos, and S.~G. Turyshev,  in {\em Lasers, Clocks and
  Drag-Free Control: Exploration of Relativistic Gravity in Space}, 
  H. Dittus, C. Laemmerzahl, and S. Turyshev, eds. (Springer Verlag, Berlin, 2007),
  pp.\ 27--67.

\bibitem{Brillet-Hall-1979}
A. {Brillet} and J.~L. {Hall}, Phys. Rev. Lett. {\bf 42},  549  (1979).

\bibitem{Hughes-etal:1960}
V.~W. {Hughes}, H.~G. {Robinson}, and V. {Beltran-Lopez}, Physical Review
  Letters {\bf 4},  342  (1960).

\bibitem{Drever-1961}
R.~W.~P. Drever, Philos. Mag. {\bf 6},  683  (1961).

\bibitem{Lamoreaux-etal:1986}
S.~K. {Lamoreaux} {\it et~al.}, Phys. Rev. Lett. {\bf 57},  3125  (1986).

\bibitem{Kostelecky:1999mr}
K.~V. Alan and L.~C. D., Phys. Rev. D {\bf 60},  116010  (1999).

\bibitem{Bell:1995ax}
J.~F. Bell, Astrophys. J. {\bf 462},  287  (1996);
%
J.~F. {Bell} and T. {Damour}, Class. Quant. Grav. {\bf 13},  3121  (1996).

\bibitem{Stecker:2001vb}
F.~W. {Stecker} and S.~L. {Glashow}, Astropart. Phys. {\bf 16},  97  (2001).

\bibitem{Kostelecky:2000hz}
V.~A. Kostelecky and R. Potting, Phys. Rev. D {\bf 63},  046007  (2001).

\bibitem{Kostelecky:1999mu}
V.~A. {Kosteleck\'y}, M. {Perry}, and R. Potting, Phys. Rev. Lett. {\bf 84},
  4541  (2000).


\bibitem{Bailey:2006fd}
Q.~G. Bailey and V.~A. Kostelecky, Phys. Rev. {\bf D74},  045001  (2006).

\bibitem{Jacobson:2000xp}
T. Jacobson and D. Mattingly, Phys. Rev. D {\bf 64},  024028  (2001);
%
B.~Z. Foster and T. Jacobson, Phys. Rev. D {\bf 73},  064015  (2006).

\bibitem{Cacciapuoti-etal:2007}
L. {Cacciapuoti} {\it et~al.}, Nucl. Phys. B (Proc. Suppl.) {\bf 166},  303
  (2007).

\bibitem{Wolf-Petit:1997}
P. {Wolf} and G. {Petit}, Phys. Rev. A {\bf 56},  4405  (1997).

\bibitem{Pound-Rebka-1960}
R.~V. Pound, G.~A. Rebka, Phys. Rev. Lett. {\bf 4},  337  (1960).

\bibitem{Pound-Snider-1964}
R.~V. Pound, J.~L. Snider, Phys.Rev.Lett. {\bf 13},  539  (1964).

\bibitem{Vessot-etal-1980}
R.~F.~C. Vessot {\it et~al.}, Phys. Rev. Lett. {\bf 45},  2081  (1980).

\bibitem{Bauch-Weyers-2002}
A. Bauch and S. Weyers, Phys. Rev. D {\bf 65},  081101  (2002).

\bibitem{Canuto-Goldman-1982-2}
V.~M. {Canuto} and I. {Goldman}, Int. Journ. Theor. Phys. {\bf 28},  1005
  (1989).

\bibitem{Uzan-2002}
J.-P. Uzan, Rev. Mod. Phys. {\bf 75},  403  (2003).

\bibitem{Dvali-Zaldarriaga-2002}
G. Dvali and M. Zaldarriaga, Phys. Rev. Lett. {\bf 88},  091303  (2002).

\bibitem{Dent-2007}
T. Dent, JCAP {\bf 0701},  013  (2007).

\bibitem{Nordtvedt-2002}
K. Nordtvedt, Int. Journ. Mod. Phys. A {\bf 17},  2711  (2002).

\bibitem{Fortier-etal-2007}
T.~M. Fortier {\it et~al.}, Phys. Rev. Lett. {\bf 98},  070801  (2007).

\bibitem{Schiller-etal-2006}
S. Schiller {\it et~al.}, Nucl. Phys. B (Proc. Suppl.) {\bf 166},  300  (2007).

\bibitem{Williams-etal-2007}
J.~G. Williams, S.~G. Turyshev, and D.~H. Boggs, Phys. Rev. Lett. {\bf 98},
  059002  (2007).

\bibitem{ADD-1998}
N. Arkani-Hamed, S. Dimopoulos, and G.~R. Dvali, Phys. Lett. B {\bf 429},  263  (1998);
%
N. Arkani-Hamed, S. Dimopoulos, and G.~R. Dvali, Phys. Rev. D {\bf 59},  086004
   (1999).

\bibitem{Adelberger-Heckel-Nelson-2003}
E.~G. Adelberger, B.~R. Heckel, and A.~E. Nelson, Ann. Rev. Nucl. Part. Sci.
  {\bf 53},  77  (2003).

\bibitem{AnDD-1998}
I. Antoniadis, S. Dimopoulos, and G.~R. Dvali, Nucl. Phys. B {\bf 516},  70
  (1998).

\bibitem{Dimopoulos-Giudice-1998}
S. Dimopoulos and G.~F. Giudice, Phys. Lett. B {\bf 379},  105  (1996).

\bibitem{Sundrum-1999}
R. Sundrum, JHEP {\bf 07},  001  (1999).

\bibitem{Dvali-etal-1998}
G. Dvali, G. Gabadadze, M. Kolanovic, and F. Nitti, Phys. Rev. D {\bf 65},
  024031  (2002).

\bibitem{Fischbach-Talmadge-1998}
E. {Fischbach} and C.~L. {Talmadge}, {\em The Search for Non-Newtonian Gravity}
  (Springer Verlag, New York, 1998).

\bibitem{Scherk}
J. Scherk, Phys. Lett. B {\bf 88},  265  (1979).

\bibitem{Bars-Visser-1986}
I. Bars and M. Visser, Phys. Rev. Lett. {\bf 57},  25  (1986).

\bibitem{Kapner:2006si}
D.~J. Kapner {\it et~al.}, Phys. Rev. Lett. {\bf 98},  021101  (2007).

\bibitem{Tu-etal-2007}
L.-C. Tu {\it et~al.}, Phys. Rev. Lett. {\bf 98},  201101  (2007).

\bibitem{Adelberger-etal-2007}
E.~G. Adelberger {\it et~al.}, Phys. Rev. Lett. {\bf 98},  131104  (2007).

\bibitem{Paik-etal-2007}
H.-J. {Paik}, M.~V. {Moody}, and D.~M. {Strayer}, Gen. Rel. Grav. {\bf 36},   523  (2004);
%
%
H.-J. {Paik}, V.~A. {Prieto}, and M.~V. {Moody}, Int. J. Mod. Phys. D {\bf 16},
   2181  (2007).

\bibitem{Chiaverini-etal-2003}
J. Chiaverini {\it et~al.}, Phys. Rev. Lett. {\bf 90},  151101  (2003).

\bibitem{pious}
J.~D. {Anderson} {\it et~al.}, Phys. Rev. Lett. {\bf 81},  2858  (1998), gr-qc/9808081;
%
J.~D. {Anderson} {\it et~al.}, Phys. Rev. D {\bf 65},  082004  (2002), gr-qc/0104064;
%
S.~G. Turyshev {\it et~al.},  in {\em Gravitational Waves and Experimental Gravity}, {\em The XVIIIth Workshop of the Rencontres de Moriond, Les Arcs, Savoi, France (January 23-30, 1999)}, edited by J. Dumarchez and J. {Tran Thanh Van} (World Publishers, Hanoi-Vietnam, 2000), pp.\ 481--486, gr-qc/9903024;
%
S.~G. {Turyshev}, M.~M. {Nieto}, and J.~D. {Anderson}, Amer. J. Phys. {\bf 73},    1033  (2005), physics/0502123.

\bibitem{pio-data}
S.~G. Turyshev {\it et~al.}, Int. J. Mod. Phys. D {\bf 15},  1  (2006), gr-qc/0512121.
%
V.~T. Toth and S.~G. Turyshev, Can. J. Phys. {\bf 84},  1063  (2006), gr-qc/0603016.
%
V.~T. Toth and S.~G. Turyshev, AIP Conf. Proc. {\bf 977},  264  (2008), arXiv:0710.2656 [gr-qc].

\bibitem{Deffayet-Dvali-Gabadadze-2002}
C. Deffayet, G.~R. Dvali, and G. Gabadadze, Phys. Rev. D {\bf 65},  044023
  (2002).

\bibitem{Deffayet-etal-2002}
C. Deffayet, G.~R. Dvali, G. Gabadadze, and A.~I. Vainshtein, Phys. Rev. D {\bf
  65},  044026  (2002).

\bibitem{Dvali-2006}
G. Dvali, New J. Phys. {\bf 8},  326  (2006).

\bibitem{Deffayet:2000uy}
C. Deffayet, Phys. Lett. B {\bf 502},  199  (2001).

\bibitem{Bekenstein:2006fi}
J. Bekenstein and J. Magueijo, Phys. Rev. D {\bf 73},  103513  (2006);
%
J. {Magueijo} and J. {Bekenstein}, Int. J. Mod. Phys. D {\bf 16},  2035
  (2007).

\bibitem{Iess-Asmar:2007}
L. {Iess}, S. {Asmar}, Int. J. Mod. Phys. D {\bf 16},  2117  (2007).

\bibitem{Spergel-etal-2006}
D.~N. Spergel {\it et~al.}, Astrophys. J. Suppl. {\bf 170},  377  (2007).

\bibitem{Ashby-Bender-2006}
N. Ashby and P. Bender,  in {\em Lasers, Clocks, and Drag-Free: Technologies
  for Future Exploration in Space and Tests of Gravity}, H. Dittus,
  C. Laemmerzahl, and S.~G. Turyshev, eds. (Springer Verlag, Berlin, 2006), p.\ 219.

\end{thebibliography}
\end{document}